\documentclass[journal=ancac3,manuscript=article]{achemso} 

\usepackage[version=4]{mhchem}
\usepackage{graphicx}
\usepackage{amssymb,amsmath}
\usepackage{color}
\usepackage{calc}
\usepackage{xspace}
\usepackage{soul} 
\usepackage{hyphenat}
\usepackage{braket}
\usepackage{caption}
\usepackage{subcaption}
\usepackage [english]{babel}
\usepackage [autostyle, english = american]{csquotes}

\author{Ayesha Amjad}
\affiliation[Department of Chemistry]
{Department of Chemistry, Indiana University, Bloomington, IN 47405, United States}

\author{Irina Tsvetkova}
\affiliation[Department of Chemistry]
{Department of Chemistry, Indiana University, Bloomington, IN 47405, United States}
\alsoaffiliation{Centre International de Formation et de Recherche Avanc\'ees en Physique, 077125 Bucharest-Magurele, Romania}

\author{Lena G. Lowry}
\affiliation[Department of Chemistry]
{Department of Chemistry, Indiana University, Bloomington, IN 47405, United States}

\author{David Gene Morgan}
\affiliation[Department of Chemistry]
{Department of Chemistry, Indiana University, Bloomington, IN 47405, United States}

\author{Roya Zandi}
\affiliation[Department of Physics and Astronomy]
{Department of Physics and Astronomy, University of California, Riverside, CA 92521, United States}

\author{Paul van der Schoot}
\affiliation[Department of Physics and Science Education]
{Department of Physics and Science Education, Eindhoven University of Technology, 5600 MB Eindhoven, The Netherlands}

\author{Bogdan Dragnea}
\affiliation[Department of Chemistry]
{Department of Chemistry, Indiana University, Bloomington, IN 47405, United States}
\alsoaffiliation{Centre International de Formation et de Recherche Avanc\'ees en Physique, 077125 Bucharest-Magurele, Romania}

\email{dragnea@iu.edu}

\title[]
  {An Assembly-Line Mechanism for \emph{In-Vitro} Encapsulation of Fragmented Cargo in Virus-Like Particles}


\begin{document}
\begin{tocentry}
\centering
\includegraphics[width = 3.25 in]{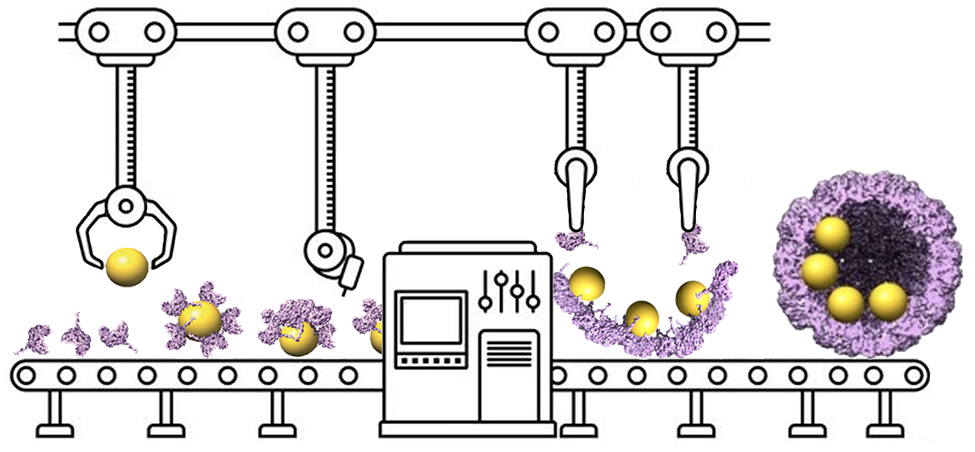}
\end{tocentry}

\section{Abstract} 

The ability of virus shells to encapsulate a wide range of functional cargoes, especially multiple cargoes - siRNAs, enzymes, and chromophores - has made them an essential tool in biotechnology for advancing drug delivery applications and developing innovative new materials. Here we present a mechanistic study of the processes and pathways that lead to multiple cargo encapsulation in the co-assembly of virus shell proteins with ligand-coated nanoparticles. Based on the structural identification of different intermediates, enabled by the contrast in electron microscopy provided by the metal nanoparticles that play the cargo role, we find that multiple cargo encapsulation occurs by self-assembly via a specific ``assembly line'' pathway that is different from previously described \emph{in vitro} assembly mechanisms of virus-like particles (VLP). The emerging model explains observations that are potentially important for delivery applications, for instance, the pronounced nanoparticle size selectivity.

\section{Introduction}

In 1950, H. Crane wrote an essay on the problem of how viruses and other biological particles replicate by association of individual molecules~\cite{Crane1950}. In the absence of molecular structure data, and equipped with few facts about the physical and chemical changes that accompany the process, Crane relied on basic principles of periodicity, automatic termination, non-self-starting, and geometry to make several predictions. Some of these predictions have been proven to be valid, and impactful. Crane wrote in his essay: ``any structure which is [...] rodlike when seen at ``low magnification'' is probably a structure having repetition along a screw axis''. Three years later, Watson and Crick~\cite{WatsonCrick1953}, revealed the double-helix structure of DNA, based on critical X-ray data by Franklin and Wilkins~\cite{FranklinGosling1953, Wilkins1953}. Crane also mentioned the key role of binding selectivity (or specificity) and suggested a way to realize that selectivity: ``... for a high degree of specificity, the contact of combining points [...] must be multiple and weak'' -- a principle that is believed to be generally responsible for the remarkable precision of certain biomolecular assemblies~\cite{oosawa1975thermodynamics, Zlotnick2003, Curk2017}. In addition, Crane posited that viral assembly does not self-start, which ensures an efficient path to completion, a principal tenet in the classical nucleation theory of virus growth~\cite{zandi2006classical}.  Finally, inspired by the ``art'' of mass production, Crane hypothesized that a multistep subassembly mechanism, similar to an assembly-line technique, could be advantageous in terms of speed and accuracy over the mechanism by which each of the final assemblies is built by adding elementary blocks individually. At least for viruses, Crane's ``assembly-line''  pathway has not been identified so far. This paper deals with an instance of cargo--protein cage \emph{in vitro} co-assembly,  featuring characteristics that in their set can be qualitatively explained by an ``assembly line'' mechanism.

Mechanistic studies of the processes leading to cargo encapsulation are useful in understanding gain or loss of function when virus-like particles are engineered for delivery of non-cognate cargo. Virus-inspired and virus-derived delivery vectors are an effective complement to other approaches in the treatment or prevention of diseases through gene and immunotherapy and vaccines~\cite{Chung2020}. Viral coat proteins are often promiscuous encapsulators, which, together with other evolution-honed features, makes them attractive as delivery vectors \cite{Tsvetkova2015, Saxena2016, Xie2021, Uchida2022, Seitz2023, Boone2024}. However, cargo formulation clearly matters for both the structure and the dynamics of the assembled complex~\cite{Zlotnick2011, bond2020virus}. The issue of how the physical properties of a non-cognate cargo affect the pathways and outcome of assembly along with the properties of the assembly result is still a very much open one. For example, when planning to engineer a delivery vector that encapsulates multiple active molecules, should the cargo be in the form of multiple, independent particles (or molecules) or single-piece, perhaps in the form of ``beads on a string'', that better resemble a nucleic acid? -- a simple question of potential technological impact that, to the best of our knowledge, has not yet been addressed directly. 

\begin{figure}
\includegraphics[width = 0.4 \textwidth]{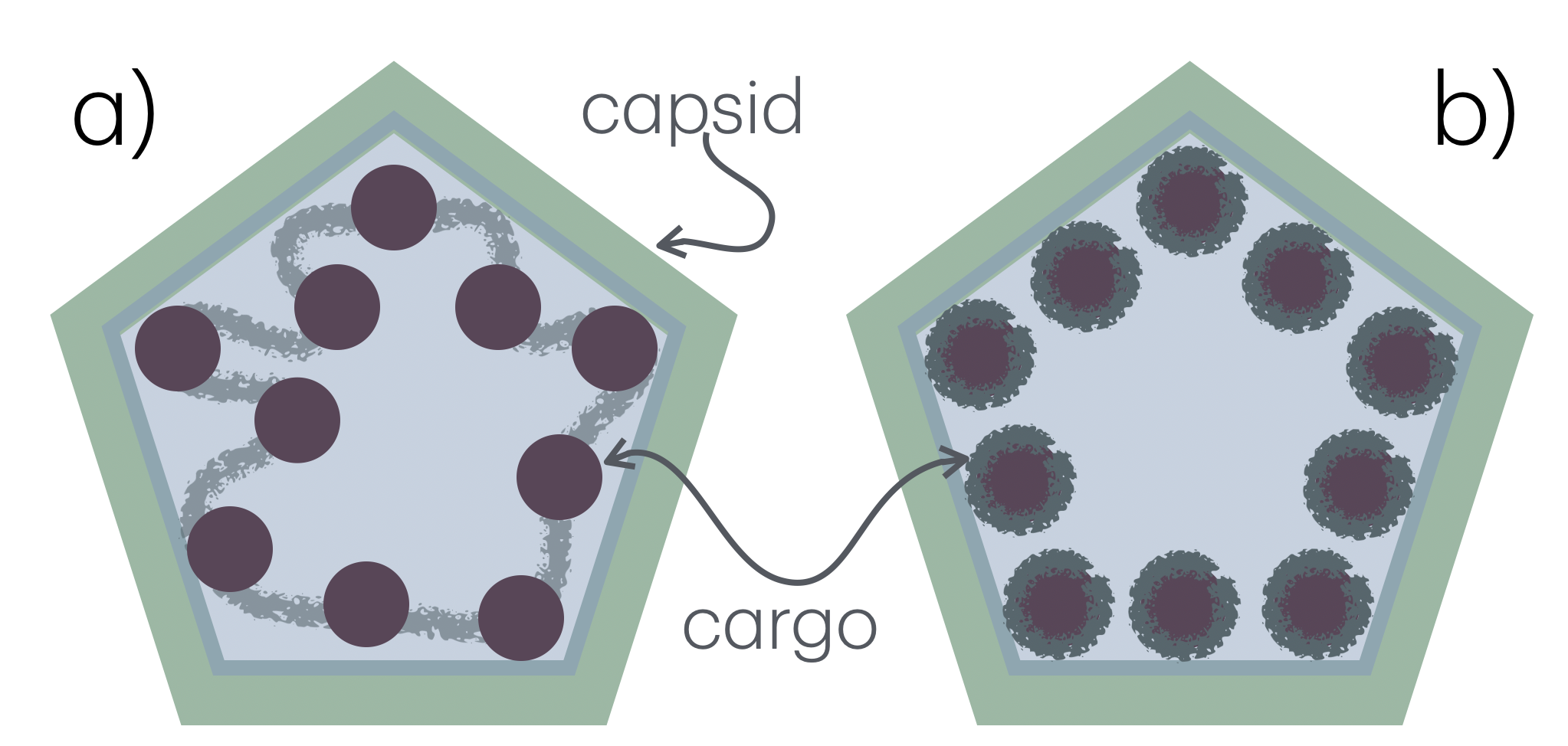}
\caption{VLPs with connected (a) \emph{vs.} fragmented (b) payload. Assuming all other things being equal, there is a larger entropic penalty to be payed at assembly in (b) and hence different thermodynamic forces driving assembly.}
\label{fig:FragmentedConnected}
\end{figure}

In the virus realm, there are examples of both fragmented and connected cargo, in the form of nucleic acid molecules sometimes in complexes with proteins, Fig.~\ref{fig:FragmentedConnected}.
The MS2 bacteriophage, for example, selectively encapsulates a single genomic RNA molecule, which is densely smeared at the lumenal interface of the capsid~\cite{Toropova2008}. The continuous strand character of MS2 RNA is considered important for selective encapsulation and genome presentation~\cite{Stockley2013}. 
In contrast, rotavirus capsids encapsulate a complete but ``fragmented'' genome comprised of several separated segments. The rotavirus nucleic acid packaging is not only selective for the correct segment sequences but also highly coordinated. The progeny virions contain one of each genome segment~\cite{McDonald2011}. 

Yet another approach to fragmented genome packaging is assumed by the brome mosaic virus (BMV). Specifically, BMV has a single-stranded RNA genome made up of four segments that are partitioned among \textit{three}, structurally identical, protein cages. This multipartite genomic organization is believed to be important for genetic assortment, replication coordination, and virion stability~\cite{noueiry2003brome,schwartz2002positive,schwartz2004alternate,miller1985synthesis}.

Recent advances have highlighted the ability of virus-derived or viromimetic protein cages to encapsulate a variety of cargoes ranging from genomes and enzymes~\cite{Worsdorfer2011, Tetter2021, Patterson2014} to molecular dyes, drugs, and nanoparticles~\cite{Guerrero2015, Le2017, chen2006nanoparticle, Benjamin2018}. As a result, virus-like particles (VLPs) are emerging as nanocarriers with great potential for enhanced functionality. For example, VLPs have been engineered to encapsulate chain-linked enzymes to facilitate cascade reactions with improved efficiency due to controlled confinement~\cite{mcneale2023tunable, sharma2017modular} or create biosynthetic organelles with restorative properties of cell metabolism~\cite{Oerlemans2021}. Furthermore, VLPs can encapsulate multiple small interfering RNAs (siRNAs)~\cite{kong2015pokemon} and scaffolding proteins~\cite{parent2005electrostatic,parent2006quantitative}, improving the precision of gene silencing and facilitating correct structural assembly~\cite{teschke2010let, fane2003mechanism}. Such capabilities expand the potential for targeted gene therapy applications. In all of these examples, the interaction between the carrier and the cargo is of great importance. 

In certain cases, the changes induced by the cargo on the molecular structure of the container cage are striking and have yielded previously unclassified cage architectures. For example, co-assembly of the BMV coat protein and multiple short oligonucleotides in VLPs was shown to lead to the formation of previously unknown strained nonicosahedral cage structures~\cite{bond2020virus, panahandeh2022virus}. This underscores the importance of understanding how the nature of cargo can affect the assembly result, as it holds the key to leveraging biological principles for synthetic biology and drug delivery applications.

In the quest to understand the interaction between cargo and assembly in the simplest viruses, which have a cage formed of multiple copies of the same protein, two mechanisms have been proposed for \emph{in vitro} assembly around a single continuous cargo. The first, known as the nucleation and growth mechanism~\cite{prevelige1993nucleation, oosawa1975thermodynamics}, was originally applied to the empty capsid assembly~\cite{zlotnick1994build, zandi2006classical} and later adapted to describe the assembly of coat proteins around generic genomic cargo~\cite{kler2012rna, hagan2009theory, kler2013scaffold}. The nucleation and growth mechanism is believed to be encountered especially when the interactions between the RNA and coat proteins are weaker than the interactions between the coat proteins~\cite{zlotnick2013build}. The mechanism was recapitulated in the assembly \emph{in vitro} of virus-like particles in which the native nucleic acid was replaced by a single nanoparticle (NP) core coated with ligand~\cite{chen2006nanoparticle,sun2007core,huang2007self,dixit2006quantum, Kuenzle2018}.

The second mechanism, initially suggested by McPherson in 2003 as an analogy with micelle formation~\cite{mcpherson2005micelle}, occurs when cargo--coat--protein interactions rapidly lead to a transient amorphous aggregate of virion components (the nucleoprotein complex), before proper rearrangement and subsequent formation of a symmetric protein cage equilibrium structure surrounding the nucleic acid cargo. In this scenario, cargo acts as a template that guides coat proteins along a restricted path to the final symmetric structure ~\cite{devkota2009structural,hagan2008controlling,elrad2010encapsulation,panahandeh2020virus}. This mechanism was observed to operate for single NP cargo through the formation of rather large aggregates from which VLPs grew~\cite{Malyutin2013} and subsequently for genomic cargo as well~\cite{Chevreuil2018}.
When genomic RNA is involved, added selectivity follows from cooperative binding to CPs, and results in simultaneous RNA condensation and capsid growth, as demonstrated by optical experiments of single capsid self-assembly in real time and MD simulations~\cite{garmann2019measurements,li2024elastomer}.

The observation of \emph{in vitro} encapsulation of multiple non-cognate cargo molecules in VLPs~\cite{Chang2008, Hu2008}, and of bacterial polyhedral microcompartments that package many enzymes at high density~\cite{Tanaka2008}, raised the question of whether the mechanisms of co-assembly of cargo and coat subunits could be different from those involving a single large cargo entity~\cite{Perlmutter2016}.  Simulations with subunits modeling the coat proteins as tiles of a predefined equilibrium shape revealed two major particle growth scenarios~\cite{Perlmutter2016}: i) growth of a cargo--protein condensate, when the cohesive interactions within the payload are strong, and ii) simultaneous association of the payload particles with a growing coat protein shell, when the cohesive interactions of the cargo are weak. 

A different computational approach adopted by Rotskoff \& Geissler avoided the predefined tiling geometry constraint, and started from the premise of a cargo species that is prone to aggregation, a shell species that has a tendency to form flat hexagonally symmetric elastic sheets, and an attractive interaction between one side of the sheet and
the cargo~\cite{Rotskoff2018}. These authors showed that under certain interaction parameters, both the cargo and the encapsulating shell could grow simultaneously under kinetic control to form a core-shell structure. All these works to date relied on assumptions of cohesive cargo-cargo interactions and of cargo elements that are much smaller than the coat subunits (\emph{i.e.} a coat subunit could bind several cargo elements, not \emph{vice versa}). However, experimental instances of encapsulation by co-assembly exist where the opposite assumptions are valid, \emph{i.e.}, the cargo is not ``finely grained'', and a single cargo subunit is insufficient to promote VLP assembly. In this case, a small number of cargo subunits must be gathered to sustain productive assembly. For example, a small number (2--4) of so-called supercharged ferritin cages loaded with iron oxide nanoparticles have been encapsulated in lumazine synthase cages by disassembly/re-assembly of the latter, to create nested, Matryoshka-type structures~\cite{Beck2015}. The outcomes of this type of assembly can be interesting: BMV coat proteins were shown to assemble \emph{in vitro} around a few single-stranded DNA oligomers to form isometric VLPs~\cite{bond2020virus}. However, the capsids of these VLPs had strained structures that were starkly different in size and symmetry from wt BMV, which assumes a Caspar-Klug canonical structure of triangulation number T = 3~\cite{panahandeh2022virus}. How such ``supercharged'' cargoes of mesoscopic dimensions with respect to the container cage get recruited to the growing particle has not yet been investigated. 

The chances are that the mechanism could be different because of two factors at least: i) charged cargo subunits strongly repel under assembly buffer conditions, and ii) by virtue of size and charge multiplicity one cargo subunit could bind several N termini of BMV CPs. These two characteristics distinguish the problem treated here from previous studies. Because VLP assembly in these specific conditions has not yet been systematically studied, yet it has biological and technological relevance, we have used BMV as a model to study the \emph{in vitro} co-assembly of CPs and small Au NPs stabilized with ligands ($1-6$ nm diameter). A single particle in this size range is too small to promote the assembly of a closed cage around itself, presumably because the elastic stress on the shell at the specific curvature imposed by the particle is too great~\cite{Moerman2016}. However, we shall see that, depending on their size and buffer conditions, charged sub-10 nm NPs can still act as promoters of assembly of BMV CP cages under conditions in which empty cages do not form, the exclusive assembly result being co-encapsulation of several NPs in a CP cage of size consistent with the native BMV one.

The benefit of working with ligand-coated metal NP cargo is that it affords visualization of both cargo and CP spatial distributions by negative stain electron microscopy (EM) and cryo-EM, at the single-particle level. As we shall see, the AFM is also able to pick-up the presence of NPs associated with capsid fragments. 

Because BMV CP shell formation around single anionic particle cores by co-assembly has been well documented both experimentally and theoretically, and its mechanisms were shown to recapitulate the two main ones discussed earlier, we have carried out experiments seeking to contrast the features of multi-NP cargo VLP co-assembly against the backdrop provided by those previous works. Our findings suggest that neither of the two previously identified mechanisms can explain all of the observations of this work. Thus, we propose an alternative mechanism where intermediate homogeneous cargo-CP complexes are formed first, and subsequently join a growing VLP, hence we describe this potential route as an ``assembly-line'' pathway, which can also be described as a multi-step nucleation and growth process, along the lines sketched by Crane. 

Unlike for the previously studied case of single, larger NP encapsulation, the diameter of multicore VLPs is not dictated by the diameter of the NP cores. Instead, the multicore VLP diameter is consistent with a T = 3 particle of 28 nm. In addition, structural characterization using cryo-electron microscopy (cryo-EM) and cryo-electron tomography (cryo-ET) shows that the particles typically cluster into patches of adjacent NPs clustered on one side rather than randomly distributed within the VLP cavity. Furthermore, there is a NP size cutoff below which NP encapsulation (and capsid formation) does not occur under set solution conditions, which confirms the prediction of van der Holst \emph{et al.} who noted in their theoretical work on the mass action law in virus assembly that there should be an entropic penalty that will work against encapsulation of increasingly small NPs~\cite{VanderHolst2018}.  Moreover, if the NP size is too small, the uncompensated charge on the protein-NP intermediate complex might lead to a net repulsive electrostatic interaction between intermediates, preventing them from joining, just as subunits without NPs cannot assemble at low salt. In summary, these are the reasons why encapsulation is NP size selective and the degree of cargo fragmentation is a control variable in encapsulation, which should require attention in both applications and predictive theoretical modeling. 

\clearpage

\section{Results and Discussion} 
BMV has served as a model system for the large Alphavirus-like superfamily of single-stranded, positive sense RNA icosahedral viruses for over 60 years~\cite{kao2000brome, Kao2011}. The BMV capsid is made up of 180 identical proteins that assume a canonical structure encountered in many small, icosahedral viruses~\cite{Casjens1985, lucas2002crystallographic}.  The primary driving forces for viral assembly are electrostatic interactions between positively charged N-termini of CPs and negatively charged RNA~\cite{kegel2004competing,ceres2002weak, zandi2020virus, belyi2006electrostatic, hu2008electrostatic, van2005electrostatics}, along with non-covalent interactions among the structural beta-barrel domains of the coat protein subunits~\cite{kegel2004competing,ceres2002weak}. The assembly of capsids \emph{in vitro} is possible at high salt in the absence of nucleic acid~\cite{krol1999rna}. 

Dragnea and co-workers developed a procedure to replace negatively charged RNA within virions with a ligand-coated gold nanoparticle (AuNP) of controlled size and shape~\cite{dragnea2003gold, Zeng2018}. This approach, adapted from the protocol for the \emph{in vitro}  assembly of cowpea chlorotic mottle virus (CCMV) around non-cognate DNA~\cite{zhao1995vitro}, showed that the co-assembly of Au NP coated with anionic ligand and BMV CP can lead to closed protein shells with structural and chemical interfacial characteristics, similar to the native virus~\cite{chen2006nanoparticle}. Specifically, work on NP-templated BMV capsid assembly was carried out with particles with diameters that were close to those of the cavity of the native virus capsid, leading to a quasi-equivalent Caspar-Klug structure similar to that of the wt virus~\cite{Caspar1962,sun2007core}. 

\begin{figure}[ht]
  \centering
 \includegraphics [width = 0.4\textwidth]{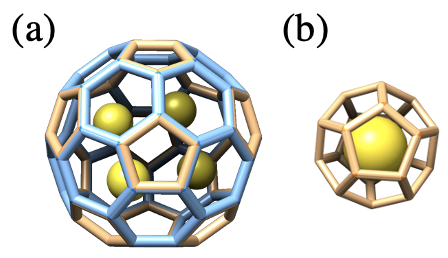}
  \caption{Chimera generated models using Cage Maker. (a) A T=3 cage containing 4 particles of 3 nm, drawn to scale within the inner cavity illustrating the type of NP-CP complex treated in this article. (b) For comparison, a T=1 cage with a 6 nm particle centered inside.}
  \label{fig:chimera}
\end{figure}

In this work, we synthesized Au NPs with diameters ranging from 1.5 to 6 nm, coated them with anionic ligands, and mixed them with BMV coat proteins under \emph{in vitro} virus assembly conditions. NPs were functionalized with a short molecular brush with ionizable end groups to provide the negative surface charge required for \emph{in vitro} assembly of wt BMV CP. An Au NP is much smaller than the internal cavity space, even for the smallest cage, T = 1, that still obeys the Caspar-Klug principles (see Fig.~\ref{fig:chimera}).  

\begin{figure}[ht]
  \centering
 \includegraphics [width = 1\textwidth]{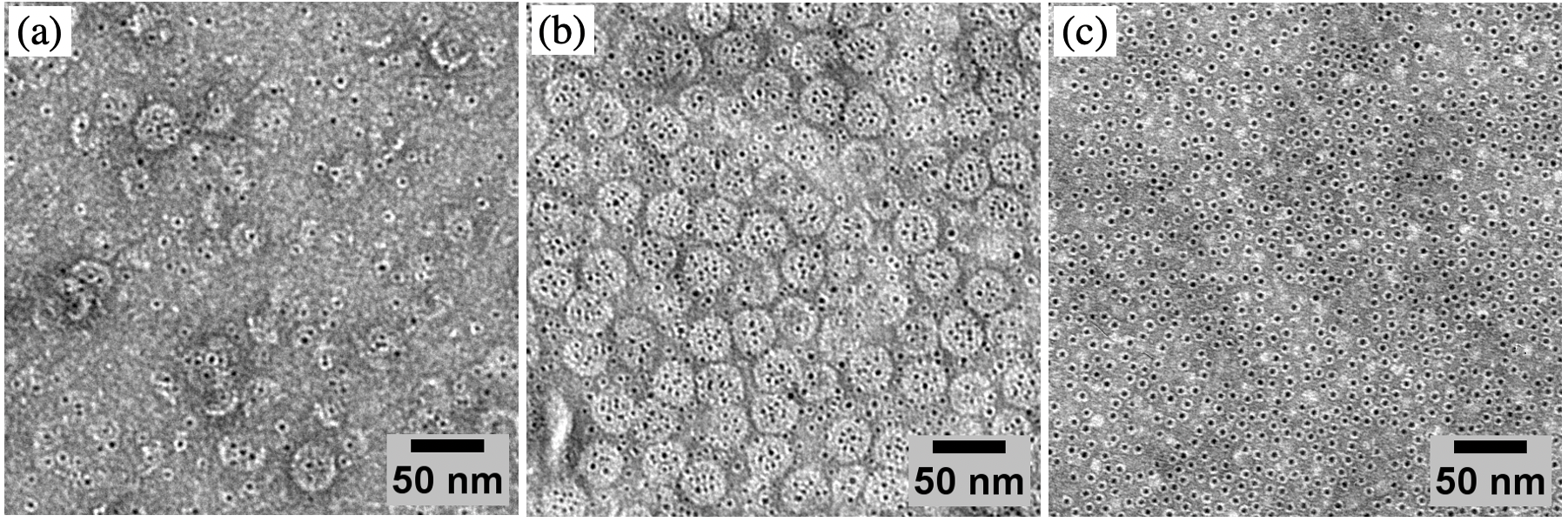}
  \caption{Negative stain TEM of the result of co-assembly of proteins and functionalized NPs at different ratios of NP/CP and fixed protein concentration (0.5 mg/ml): (a) 3:180, (b) 7:180, and (c) 10:180.}
  \label{fig:Diffratios}
\end{figure}

The starting point in the study of co-assembly of small Au NPs and BMV CPs was the same mixture conditions that led to efficient assembly of single NP core BMV VLPs~\cite{daniel2010role}, that is, a protein-to-particle ratio of 180:1 - sufficient in principle to form a T=3 capsid. However, under the same conditions, while the protein concentration was kept constant, we observed aggregation of particles and proteins but no VLP assembly over a time of two days. This observation aligns with the entropy considerations discussed by van der Holst et al.~\cite{VanderHolst2018}, where insufficient driving forces for ordered assembly can lead to aggregation rather than VLP formation. This suggests that increasing the concentration of NPs might eventually reach a critical threshold for successful nucleation and growth. We then increased the NP concentration in the assembly mixture, specifically to 180:3 (CP:NP), 180:7, and 180:10 proteins. At 180:3, the assembly reaction yielded mainly partially assembled complexes (Fig.~\ref{fig:Diffratios}a). 

\begin{figure}[ht]
  \centering
 \includegraphics [width = 0.5\textwidth]{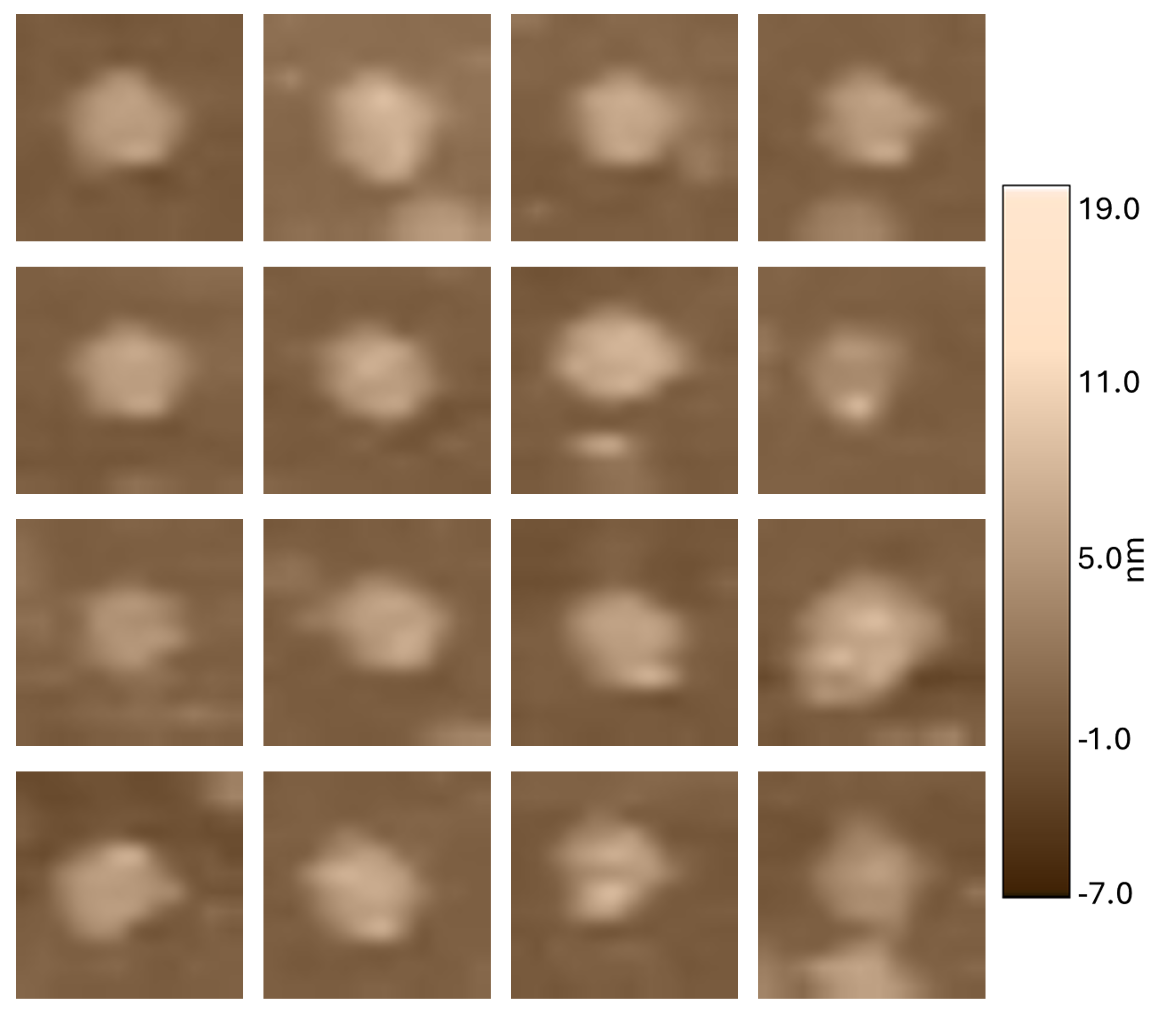}
  \caption{Array of AFM close-ups, $47 \times 47 $ nm,  showing surface-adsorbed capsomeric shapes from samples that formed only partials at self-assembly.}
  \label{f:zoomin}
\end{figure}

Partial capsids readily disassembled as they adsorbed on a surface, as we found out by attempting to image them by AFM, in assembly buffer (supporting information, Fig.~S2). The fragment sizes were remarkably homogeneous, Fig.~S2, suggesting a structural oligomer. Their height and width are consistent with those of a pentamer or hexamer of dimers after tip size deconvolution. Moreover, some of the adsorbed capsomers often displayed one or two bumps of height consistent with that of NPs, Fig.~\ref{f:zoomin}. It should be emphasized that, in absence of NPs, such CP oligomers were not observed.

\begin{figure}[ht]
  \centering
 \includegraphics [width = 0.6\textwidth]{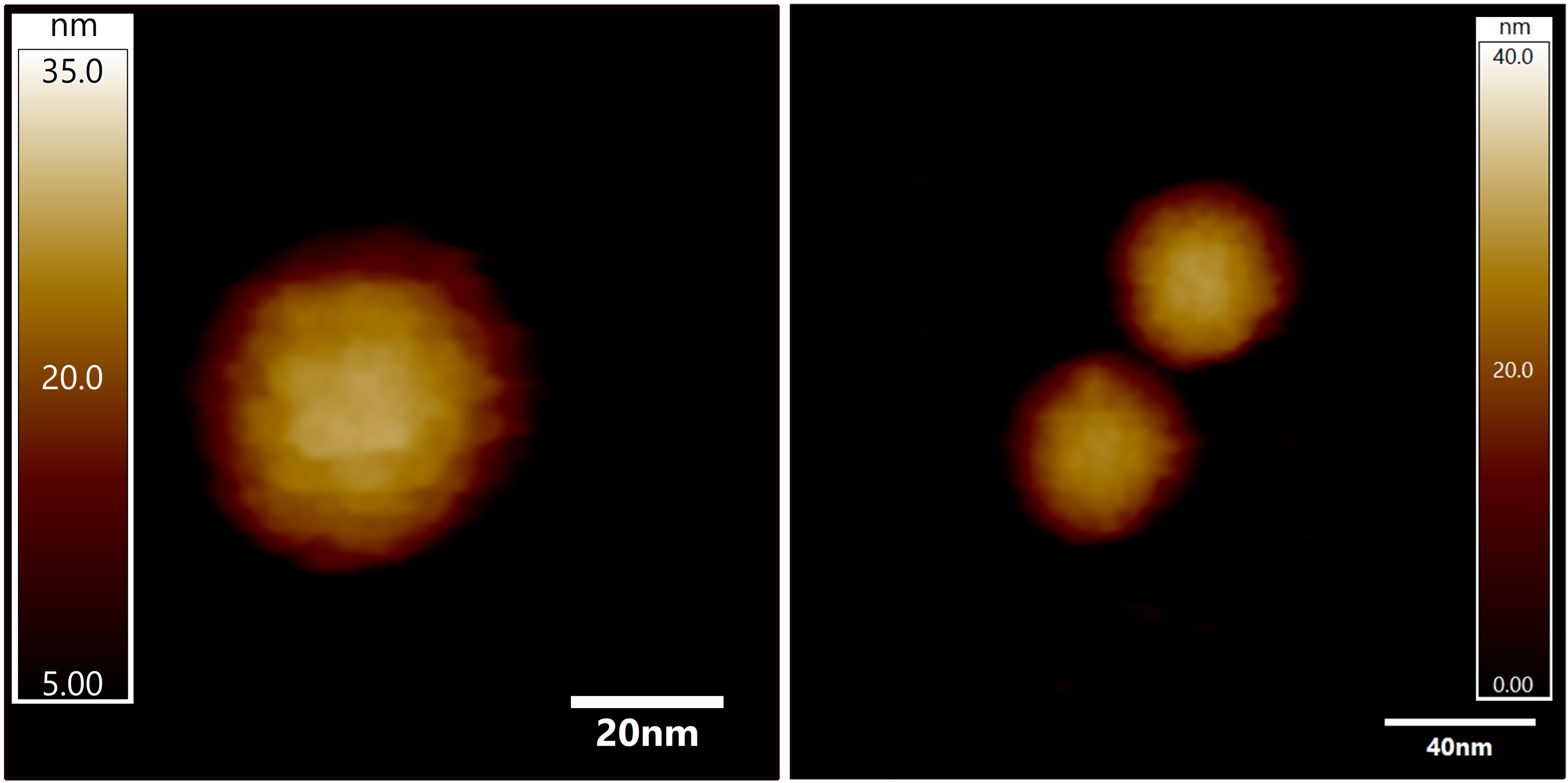}
  \caption{AFM images of complete multi-core VLPs. A canonical T=3 arrangement of capsomers can be observed in the image on the left. The micrograph on the right is lower magnification and shows particles with capsomer patterns and having a diameter consistent with T=3.}
  \label{f:T3}
\end{figure}

Assembly of complete VLPs occurred at 180:7 ratio (Fig,~\ref{fig:Diffratios}b). Microscopic examination of the assembled VLPs found that the VLPs contained multiple NP-cores and no single-core VLPs were formed. The arrangement of CPs in the shell was consistent with a T=3, Caspar-Klug structure, Fig~\ref{f:T3}.

At 180:10 proteins, no VLPs were formed; instead, ligand-coated free NPs, sometimes associated to CP clusters, could be observed (Fig,~\ref{fig:Diffratios}c). 

Control experiments conducted without the addition of gold NPs under the same assembly buffer conditions and CP concentrations revealed that no empty virus capsids were formed, suggesting that capsid assembly is induced by the presence of gold nanoparticles. 

\begin{figure}[ht]
  \centering
 \includegraphics [width = 1\textwidth]{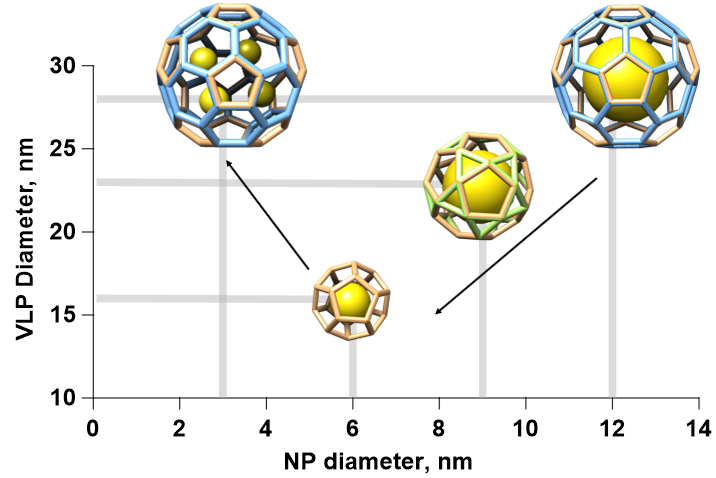}
  \caption{Non-monotonical trend in VLP diameter \emph{vs.} NP diameter from previous and current work. When NP diameter drops below $\sim 6$ nm, VLPs of diameter consistent with a T=3 capsid assemble around multiple NP cores. Above the 6 nm NP size threshold, a single encapsulated NP core is observed inside a protein cage. In this case, the cage diameter changes proportionally to the diameter of the cargo NP, as observed in previous instances~\cite{sun2007core, He2013}}
  \label{fig:scheme_VLPs}
    \end{figure}

\begin{figure}[ht]
  \centering
  \includegraphics[scale=0.48]{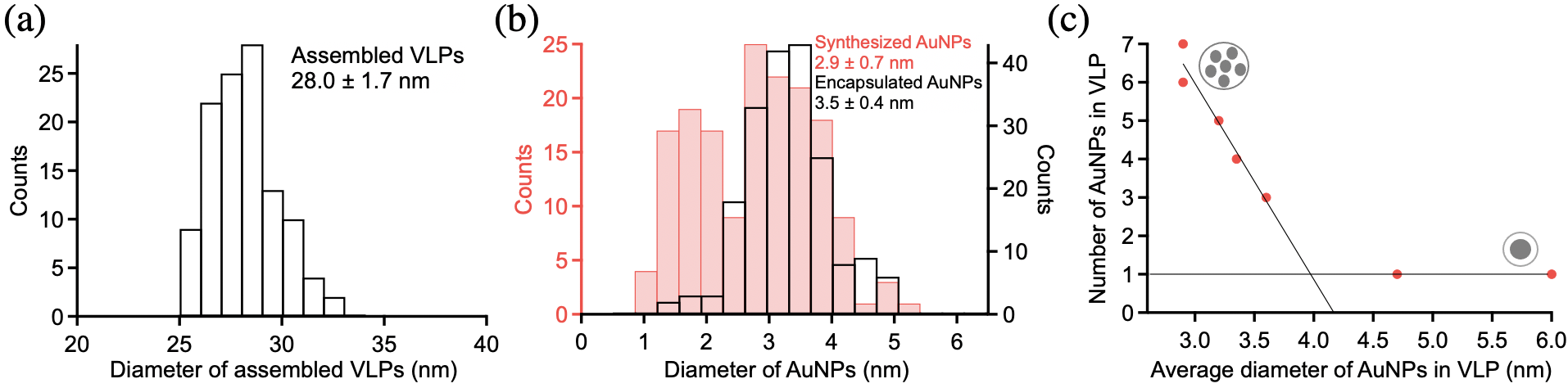}
  \caption{Size and number analysis of the assembled VLPs from the negative stain TEM (a) Diameter of the multi-core VLPs has an average diameter of 28.0 ± 1.7 nm, consistent with a native BMV diameter. (b) NP size selectivity of encapsulation is observed when a broad distribution of Au-TEG nanoparticles are added in the assembly mixture. The histogram in red represents the diameter of the synthesized Au-TEG NPs, while the histogram in black corresponds to the diameter of the encapsulated Au-TEG NPs. (c) The number of Au-TEG NPs encapsulated within the assembled multi-core VLPs is strongly influenced by the average NP diameter. There is a critical threshold of approximately 4.5 nm where single core, T=1 VLPs, start forming.} 
  \label{f:-vestain_size}
\end{figure}

Earlier studies on the formation of VLPs around single NP cores have demonstrated that the diameter and structure of the VLP shells are influenced by the encapsulated NP~\cite{sun2007core, He2013}. Specifically for Au-BMV VLPs, as the diameter of NP decreased from 12 nm to 9 nm, so did the number of proteins in the shell, going through progressively smaller T numbers, from 180 (T=3) to 60 (T=1) (Fig.~\ref{fig:scheme_VLPs}). Intriguingly, this work shows that the size of the VLPs encapsulating NPs smaller than 6 nm is 28 nm, which is consistent with a T = 3 BMV shell (Fig.~\ref{fig:scheme_VLPs}, Fig.~\ref{f:-vestain_size}a). This result strongly deviates from the previously observed trend, where the decrease in the single-core size led to smaller VLPs. 


We found that, when a NP sample with a relatively broad size distribution all below $\sim 5.5$ nm is mixed with CPs under co-assembly conditions, not all NP diameters are encapsulated.
The variance of encapsulated NP diameters is reduced with respect to that of the initial NP solution and the mean size of the encapsulated NPs is increased with respect to the initial NP size distribution (Fig,~\ref{f:-vestain_size}b). The reason for the reduced variance in the size distribution of the encapsulated particles is that NPs with diameters below $\sim 2.2$ nm were excluded from assembly. To ensure that this is not an artifact caused by insufficient negative stain contrast from smaller particles, we performed TEM imaging with and without staining. In the latter case, NP contrast dominates over imaging noise even for the smallest particles. This confirmed a truncation in the NP size distribution occurring at small diameters, which suggests that a critical NP diameter is required for encapsulation.

The number of nanoparticles per VLP was extracted from the TEM images. As stated, the diameters of the VLPs were all around 28 nm. The minimum number of encapsulated particles was 3, the maximum was 7, with an average of 5 Au NPs per VLP (Fig.~\ref{f:-vestain_size}c). As expected, the number of encapsulated NPs is inversely proportional to the average diameter of particles within the VLP.  Furthermore, when the average diameter of the NPs increases sufficiently, the formation of single core VLPs with shell sizes consistent with a structure of T = 1 is observed (supporting information, Fig.~S1). Thus, a crossover between multi-core and single-core encapsulation is observed at $\sim 4.5$ nm NP diameter (Fig.~\ref{f:-vestain_size} c). At the crossover, the size of the VLPs changes (decreases) to $\approx19$ nm, which is consistent with a T=1 particle and previous studies.

It is generally accepted that co-assembly of anionic cargo with CP under conditions of lower ionic strength (than those conducive to empty capsid formation) is primarily driven by electrostatic interactions. In viruses, the charge ratio between the anionic cargo and the protein N-termini is typically 1 to 1.6, that is, overcharging is typically observed~\cite{belyi2006electrostatic, hu2008electrostatic}. Charge regulation, \emph{i.e.}, the dependence of the net charge of ionizable groups on the surrounding chemical conditions~\cite{kusters2015role}, also plays a role. However, based on no assembly at high ionic strength, if we assume near charge neutrality as a factor for multicore VLP formation, the minimum total charge required to form a T=1 capsid would be 540 equivalents $e^{-}$, while for a T = 3 capsid it would be 1620 $e^{-}$. The estimated charge on a 2.9 nm TEG-coated NP is 185 $e^{-}$ and that on a 3.7 nm NP is 269 $e^{-}$,~\cite{daniel2010role} both of which are below the charge threshold required to form a single-core T=1 or a T=3 capsid. Therefore, multiple NPs are necessary to provide enough charge to stabilize the entire capsid against electrostatic repulsion between charged N-termini.

\begin{figure}[ht]
  \centering
  \includegraphics [width = 1\textwidth]{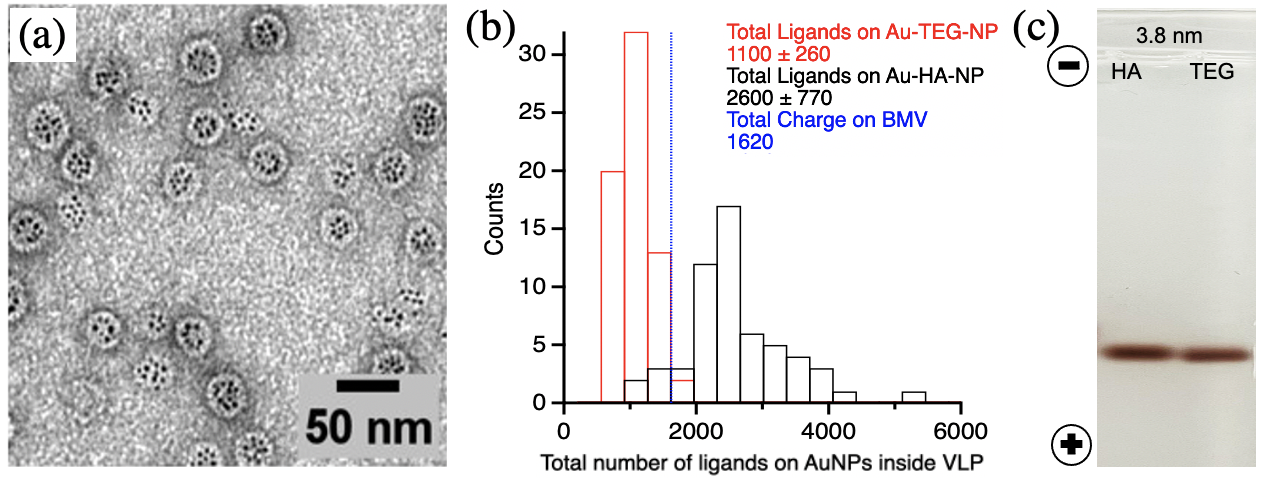}
  \caption{(a) Negative stain TEM of Au-HA NPs encapsulated VLPs. (b) Comparison of VLPs assembled around TEGylated NPs and HA NPs. The average number of ligands in VLPs with TEGylated NPs was $1100 \pm 260$, while that with the hexanoic acid ligands was $2600 \pm 770$. The total positive charge on the BMV interior is 1620. (c) Gel electrophoresis of Au-TEGylated and Au-hexanoic acid NPs indicates very similar mobility.}
  \label{f:Charge}
\end{figure}

Under the premise that total charge is an important determinant for packaging, since encapsulated NPs have size as well as charge variance~\cite{kusters2015role}, we obtained the total NP surface area from the number and diameters of NPs for a large number of VLPs imaged by TEM, Fig.~\ref{f:Charge}a. Knowing the total area of encapsulated NPs, we calculated the mean total number of ligands coating encapsidated NPs based on previously obtained ligand density estimates~\cite{daniel2010role, Xia2012}. Fig.~\ref{f:Charge}b presents the histogram of estimated total number of ligands carried by encapsidated Au-TEG-NP and Au-HA-NP along with the number of charges on the BMV capsid N-termini. While the total positive charge associated to the interior of the T=3 virus capsid is 1620 $e^{-}$, the maximum total surface charge carried by NPs is between 900 $e^{-}$to and 2000 $e^{-}$ for TEG-NPs, while for HA-NPs the total maximum charge is between 1000 and 4000. Interestingly, the most frequent total number of ligands coating encapsidated multi-particle cargo is $1100 \pm 260$ for TEG, while for HA is more than double: $2600 \pm 770$. We also note that instances of assembly were observed in both cases where the total surface charge of NP was well below the neutralization value of the capsid N-termini, which appears to be unlike the encapsulation of multiple soft polyanions~\cite{garmann2016physical, bond2020virus} but has been predicted in previous theoretical work on single particle encapsulation~\cite{kusters2015role}. It is also worth noting here that heterogeneity in the cargo charge in co-assembly is likely very high, based on the number of ligands on encapsidated particles. Cargo charge heterogeneity was theoretically predicted to significantly alter cooperativity and therefore, along with undercharging, the mechanism of encapsulation~\cite{lin2012impact}.

A question that arises in regard to the experimental size distributions of encapsidated NPs is why particles below 2 nm could not be observed in multi-core VLPs when at comparable concentrations as larger NPs? To answer this, we may focus on the chemical equilibrium between assembled and disassembled states
\[
\nu_r P_{r} + \nu_{CP} CP \rightleftharpoons VLP_r
\]
where $\mathrm{\nu_r}$ is the stoichiometric coefficient of particle types $r=1,2$ with different radius, $\mathrm{P_r}$, and $\mathrm{\nu_{CP} = 180}$ is the stoichiometric coefficient for the coat protein, CP. The ratio of the equilibrium constants $K_r$ for two different particle types is then:
\begin{equation}\label{eq:lawofmassaction}
    \frac{K_1}{K_2} = \frac{X_1 x_1^{\nu_2}}{X_2x_2^{\nu_1}} 
\end{equation}
where $x_r$ denotes the mole fractions of the particle type $r$ and $X_r$ that of the VLP encapsulating that particle type, where in dilute solution $x,X\ll1$. Note that the mole fraction of CPs cancels out. Even if the two particle types produce the same free energy gain in the VLPs, and $K_1=K_2$, we conclude that
\begin{equation}
   \frac{X_1}{X_2} = \frac{x_1^{\nu_1}}{x_2^{\nu_2}} \approx x^{\nu_1-\nu_2}, 
\end{equation}\label{eq:stabilityratiovlps}
where $\nu_1>\nu_2$ for the smaller type nanoparticle if $x = x_1\approx x_2$. 
So, the VLPs co-assembled with the smaller type of particle are much less likely to form compared to those with the larger type. This is caused by an increasing entropy penalty associated with encapsulation as the diameter of the NP decreases and the number of encapsulated particles increases~\cite{VanderHolst2018}. In follow-up work, we delve deeper into statistics of multi-core encapsulation and the role of electrostatics therein~\cite{vdSchoot2025}, extending the earlier work of van der Holst \emph{et al.}~\cite{VanderHolst2018}.

We do not need this kind of extensive modeling to be able to conclude from Fig.~\ref{f:-vestain_size} that the number of smaller particles required to encapsulate in order to obtain a stable particle is inversely proportional to the diameter of the NPs. From eq.~\ref{eq:lawofmassaction} we conclude that this suggests that $K_1 > K_2$ defining particle 1 to be larger than particle 2 and $\nu_1 < \nu_2$, if we put $X_1\approx X_2$. In other words, larger particles not only lose less translational entropy but in addition have a larger binding energy than smaller ones do.
Examining the electrostatic contribution to encapsulation, NPs with average diameters of 3.7 nm provide enough charge to stabilize approximately 16 dimers (free CP exists in dimer form in solution). This is consistent with 3 pentamers of dimers or 2 hexamers of dimers, \emph{i.e.}, 2 or 3 capsomers per NP. Notably, for CCMV -- a close relative of BMV, \emph{in-vitro} experiments at high salt conditions suggested the formation of a critical nucleus~\cite{prevelige1993nucleation} for empty capsids to be a pentamer or a hexamer of dimers~\cite{zlotnick1994build}. Thus, assuming that the CP oligomer intermediate serving as a nucleus in BMV is similar in the nucleation-and-growth mechanism for VLPs, the 3.7 nm NPs are likely to carry enough charge to condense sufficient CP for a nucleus CP oligomer. By contrast, NPs with diameters between 1--2 nm can carry at most 50 $e^{-}$ charges (if all ligands are charged). However, due to proximity, screening, curvature, and charge regulation effects, the fraction of charged ligands can be as small as 50\%~\cite{kusters2015role, Solveyra2016}. Therefore, the surface charge carried by NPs smaller than 2 nm is only sufficient to neutralize the charges on the N-termini of only about 1 or 2 CP dimers. In addition, only a fraction of the surface charge is seen by the arginine-rich motifs on the N-termini since the Debye length is about 1 nm only. These observations suggest that the smallest particles may not be able to form the nucleus CP oligomer. However, the cut-off point in the diameter of the encapsulated NP in our broad NP size distribution experiments suggests that even if the growth could be started by a 3.5-nm NP, the encapsulation of 2-nm particles will not occur at any stage of growth. Further growth also seems to require NPs that are larger than 2 nm and are bound to CP oligomers. This suggests that growth occurs by addition of CP-NP oligomers or CPs.  

Several effects are likely contributing to the observed cut-off between single-core and multi-core assembly regimes. One is steric hindrance: two 3.5 nm NPs cannot fit into a $T=1$ capsid. Another is entropy: several NPs lose more translational entropy than one. The third is the interplay between the electrostatic energy and elastic energy excess: previous findings showed that single-core VLP assembly does not occur unless a critical charge density on NPs of approximately 2.8 e /$\mathrm{nm}^{2}$ is met~\cite{daniel2010role}. Since the diameter of the NP was fixed in these experiments (at 6 nm for a T=1 shell), and given the estimated density of the ligand and the surface area of the particles in the previous work, the total maximum charge leading to a T = 1 VLP containing a single 6 nm NP was $\approx 360$e$^-$.
At this surface charge, the electrostatic interaction energy must have been high enough to overcome the barrier of elastic deformation away from the preferred radius required to form a T = 1 particle (instead of T = 3)~\cite{Timmermans2022}. In contrast, the charge density on a 3.5 nm NP is $\sim$2 e/nm$^2$ which is slightly below the charge density reported for T = 1 VLP assembly. We hypothesize that in our case, due to the smaller diameter of NPs, which implies much larger elastic stresses than in a wt shell, the energetic barrier for a T = 1 assembly could not be overcome. Presumably, the elastic energy barrier to the formation of a shell consistent with a T = 3 capsid being minimal, formation of a T = 3 capsid around a sparse set of NPs is favorable under buffer assembly conditions. We note that even for a shell of optimal radius there must be a nucleation barrier, whose height depends on the radius associated with an incomplete intermediate shell~\cite{zandi2006classical}.

Because the NP surface charge density is a function of the ligand length at a fixed surface bond density, we attempted to obtain independent control of the NP surface charge by using a ligand with a shorter length but the same type of covalent surface bond.  We chose mercaptohexanoic acid (HA) which is almost three times shorter in length than TEG. However, we note that, because of the chemical differences between HA and TEG, this approach has the caveat that different intermolecular interactions between the ligand and the protein could also lead to differences in the assembly mechanism. With respect to the great difference in length, chemical differences will be considered a perturbation.
Assuming that all ligands are charged and the density of the surface sites covalently bound to the ligand is the same, the maximum surface charge density for NPs coated with HA ligands would be 4 e/nm$^2$, which is twice the maximum charge density of TEGylated NPs for the same radius of NP.

The co-assembly of CP with Au HA-NP occurred under conditions similar to those for TEG-NP co-assembly. Similarly to TEG-NPs, no single core VLPs were observed for HA-NPs co-assembly (Fig.~\ref{f:Charge}a).  Instead, VLPs contained an average of eight HA-coated NPs, which is $ \sim$ 50\% more than the average number of Au-TEG NPs per VLP (Fig.~\ref{f:Charge}a). The same protocol as the one used for TEG-NPs led to Au-HA NP VLPs of size consistent with a T = 3 particle, but the assembly results also included numerous malformed complexes (for TEM images, see Supporting Information, Fig. S3). To enhance the yield of properly assembled VLPs, we cycled the pH of the solution between the assembly and disassembly conditions. The idea was that the most stable particles would be preserved, while the malformed and incomplete shells would disassemble. Then free CPs would have the opportunity to co-assemble with available NPs during the next cycle. The assembly efficiency did improve and the new protocol produced a more homogeneous sample made up of a majority of multi-core VLPs (Fig.~\ref{f:Charge}a). This deviation from the TEG protocol highlights differences in assembly kinetics between the different ligands. At equilibrium, the same capsid structure was observed for Au-HA or Au-TEG NPs, but the VLPs contained a different average number of NPs clearly indicating a role for the NP interfacial properties distinct from just size.

To explore whether the observed differences in assembly may be due to total charge differences between the two ligand species, we performed gel electrophoresis and zeta potential measurements on 3.8 nm NPs coated with TEG and HA ligands, Fig.~\ref{f:Charge}c. The measurements of the zeta potential indicated that both ligands had a zeta potential of -39 $\pm$ 5 mV. This is not surprising because a charge-regulated, natural surface potential must be about 25 mV, which corresponds to a surface Coulomb energy of 1 $k_BT$. If larger, either counterion condensation or deionisation of chargeable groups will occur. 

Within the experimental detection limits, both ligands exhibited bands at the same position on the gel with the HA-NPs migrating slightly behind TEG-NPs. Therefore, electrokinetic measurements do not clearly corroborate the differences between HA-NPs and TEG-NPs in the total encapsidated NP surface area. The discrepancy could come i) from the complex relationship between surface charge and electrophoretic mobility, ii) from the dominance of other intermolecular phenomena than charge interactions at assembly, for instance, contributions from the conformational entropy of the functionalized tethers might be different -- this depends on their length.  Even in the brush regime, charging up the chains may change the projected length of the tethers. For short ones this is less possible, and for longer ones as they have less ``stored'' length to offer, and iii) because the charged state of particles in free solution and those encapsulated may not be the same. Specifically, interacting with the basic N-termini is expected to increase the charged fraction of the ligands~\cite{kusters2015role}. 

Gel electrophoresis is a complex phenomenon that requires elaborate numerical modeling for quantitative predictions~\cite{Tsai2011}. However, a basic theory formulated more than a century ago by Henry~\cite{henry1931cataphoresis} correctly identifies the principal determinants of mobility and their qualitative influence.  For example, mobility will increase, in general, with $\kappa a$ -- the ratio between the NP radius ($a$) and the Debye length, $\kappa^{-1} = \sqrt{k_B T \epsilon_s \epsilon_0/(2 I e^2)}$, where $k_B$ is the Boltzmann constant, $T$ is the temperature, $\epsilon_s$ is the electrical permittivity of the solution, $\epsilon_0$ is the vacuum permittivity, $I$ is the ionic strength (in units of $\mathrm{m^{-3}}$), $e$ is the elementary charge. However, for another important factor, the zeta potential ($\zeta_a$) to correct the trends requires higher-level theories, which predict that mobility in $\kappa a \lesssim 1$ is directly proportional to $\zeta_a$, while mobility at $\kappa a \gtrsim 1$ will decrease with $\zeta_a$~\cite{Tsai2011}. Charge regulation adds to the complexity of the problem.  It is possible that our experimental conditions coincide to form this crossover. Unfortunately, significantly changing the ionic strength or the NP radius is not suitable to test this hypothesis.  

Another factor that may play a role in the similarity of the electrokinetic results between the TEG and HA NPs is the stronger ligand-ligand Coulombic interaction in the latter due to a smaller radius of the charge shell, which could lead to regulation of the ligand charge towards the protonated state in HA, and therefore a decrease in the total charge in HA-NP with respect to the maximum expected according to the ligand density~\cite{wang2011and, nayak2024ionic}. 

Regardless of the explanation for the similarity in electrokinetic properties, we conclude that the significantly different number of encapsidated NPs between the two ligands is probably due mainly to the fact that TEG NPs have a diameter larger than that of HA NPs, with fewer of them being encapsulated within the virus capsid compared to those of HA NPs. However, charge-density effects should not be excluded based on the similarity of the electrophoretic data (see discussion in Supporting Information, Section S1).
 
A final difference between the two ligands is the assembly efficiency for HA, which was lower than that for TEG-NPs under the same conditions, despite the similar electrophoretic mobility. However, here the size difference cannot explain the regaining of assembly efficiency by the cycling process. The cause is more likely to be a difference in the intermolecular interactions between CP ligands. We note that pKa inside and outside the capsid is likely different and that slight differences in pKa can have a large effect. Moreover, the more hydrophobic character of the hexanoic acid ligand and the antifouling properties of TEG-containing ligands that are more ``slippery'' to proteins~\cite{ostuni2001survey,you2006engineering} are likely to be a factor. It is not unreasonable to assume that TEGylated ligands allow protein dimers to diffuse more easily on the NP surface and to optimally orient themselves as the ionic strength decreases, while a stronger interaction between proteins and HA ligands may reduce the lateral diffusion of protein dimers, hindering their rearrangement and leading to kinetically trapped assemblies.

Negative stain TEM data, while valuable in many ways, have limitations in terms of extracting three-dimensional (3D) structural information.  TEM provides projection images under high vacuum conditions, and therefore, soft hydrated samples are desiccated and partially altered structurally. At the same time, in the words of D'Arcy Thompson, ``biological shape is a diagram of forces'' -- the actual 3D VLP structure possibly contains information revealing the mechanism of multicore VLP formation. As a related example, the morphology of colloidal lattices on curved surfaces highlights the interplay of geometry, elasticity, attractive interactions~\cite{meng2014elastic}, and topological defects~\cite{paquay2016energetically} in assembly mechanisms. Therefore, we set out to obtain 3D structures of VLPs containing multiple NP cores.

\begin{figure}[ht]
  \centering
 \includegraphics [width = 1 \textwidth]{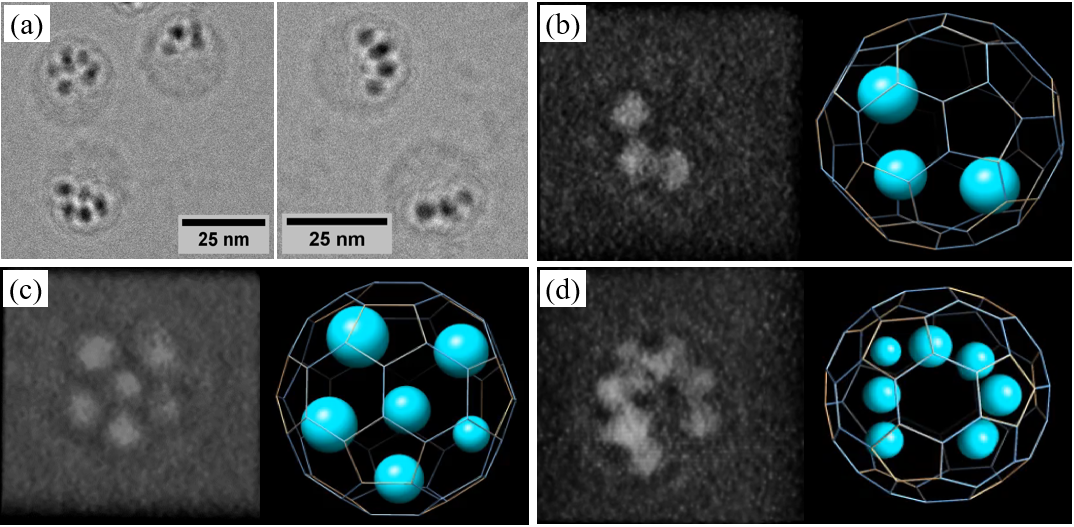}
  \caption{Typical structures of multicore (TEG-NP) VLPs  (a) Cryo-EM projections of the multicore VLPs showing partial occupation of the capsid lumenal cavity. (b--d) Reconstructed tomograms, along with the 3D locations and diameters of the encapsidated NPs (turquoise), The orientation and the structure of the icosahedral cage is arbitrary; it provides a guide the eye for the location of the CP shell outer surface.}
  \label{fig:Cryo}
\end{figure}

Individual particle architectures were obtained by tomographic reconstruction from cryo-EM micrographs (Supporting Information, Fig. S4). These images were used for tomographic reconstruction. For both HA- and TEG-NP VLPs we observed only $\sim20$\% of VLPs to be completely filled with AuNPs. A large fraction of the VLPs ($\sim$70\%) were partially filled with NPs.  The tomography of partially filled VLPs revealed that NPs are placed in adjacent locations, clustered on one side of the VLP, indicating the existence of attractive interactions between capsid-bound NPs, Fig.~\ref{fig:Cryo}.
Encapsulated NPs are organized into linear arrays or spherical caps that line the inner surface of the CP shell. Tomogram videos can be found in the Supporting information (Supporting Videos, see S1-S3). 

The adjacency of encapsidated NPs suggests that there is an attractive interaction between the polyanionic cores. Because free NPs in solution repel under assembly conditions, the attractive interaction must be provided by the CP N-termini. On the basis of charge neutrality requirement, which we have seen has limitations related to overcharging and undercharging, roughly two capsomers can be electrostatically bound to an NP, but it is also possible that some N-termini bridge between adjacent NPs. It follows that capsomer-bound NPs can be associated upon encapsulation into rafts that have to adapt to the spherical geometry of the lumenal cavity wall. An energetic frustration can then arise from the growth process that prevents complete filling. This situation is reminiscent of the strained colloidal growth of spherical crystals observed by Manoharan \emph{et al.}~\cite{meng2014elastic}. It is possible that, similar to their experiments, we are dealing here with a case of excess elastic energy that arises as a collective effect in a NP array connected via CP N-termini.

We can now summarize characteristics that are perceived as uniquely present in multi-core VLP assembly and attempt to suggest a mechanism based on them:
\begin{enumerate}
    \item Adjacency. Encapsidated NPs are organized as contiguous linear or spherical cap arrays that line the VLP cavity.
    \item Size threshold. There is a critical NP size ($\gtrsim 2$ nm, presumably concentration dependent) that has to be exceeded for encapsidation at any stage.
    \item Undercharging. Multi-core VLP growth and completion can occur even when the charge carried by NP cores is less than the amount required to neutralize the charge on the lumenal capsid interface. However, empty capsids are not observed.
    \item Narrow CP:NP window. VLPs are the dominant result from co-assembly of CP and small NPs  only in a narrow window of CP:NP molar ratios. 
    \item Unlike small flexible polyanionic nucleic acid oligomers, which lead to the formation of small strained capsids~\cite{bond2020virus}, small rigid nanoparticles assemble into multicore VLP structures consistent with a T = 3 BMV capsid. This could be due to the volume occupied by nanoparticles compared to nucleic acid oligomers.
\end{enumerate}

Because empty capsids are not observed at the concentrations in this paper, the first step requires the association of several CP dimers with a NP. A pentamer of dimers will have roughly an equal number of positive charges as the negative surface charge estimated for an NP. For the concentration range studied here, when the NP:CP ratio is higher than $\sim 10:180$ (and the mean NP size is larger than 2 nm) the process stops there. No VLPs will form under the given conditions, see Fig.~\ref{fig:Diffratios} c). In Fig.~\ref{fig:Diffratios} a) we see some incomplete or malformed shells attempting to grow from a single CP-NP or a small cluster of NPs. Therefore, after the initial CP-NP binding, particle growth is mostly mediated by the coalescence of weakly charged, high surface-area NP-capsomer heteromers (and less likely by addition of CP alone). We note that the large intermediate incomplete capsid structures are reminiscent of the bowl-like structures observed in the case of P22 assembly in the presence of scaffolding proteins~\cite{parent2005electrostatic}. 

For successful VLP formation, Fig.~\ref{fig:Diffratios} b), we posit that a sufficient number of CP-NP heteromers must occur at intermediate NP:CP ratios. These CP-NP heteromers are weakly charged and therefore can associate into larger structures. In this preformed CP-NP intermediates scenario, the interesting characteristic of adjacency and partial emptiness of the lumenal cavity can be explained as the reflection of three steps: i) initial association formation of NP:CP heteromer intermediates, ii) formation of NP:capsomer heteromers, and iii) closure of the capsid, after a rim energy barrier~\cite{zandi2006classical,Luque2012} has been exceeded, which can be achieved by adding free CP, hence the ``empty'' caps in VLPs.

\begin{figure}[ht]
  \centering
 \includegraphics [width = 1\textwidth]{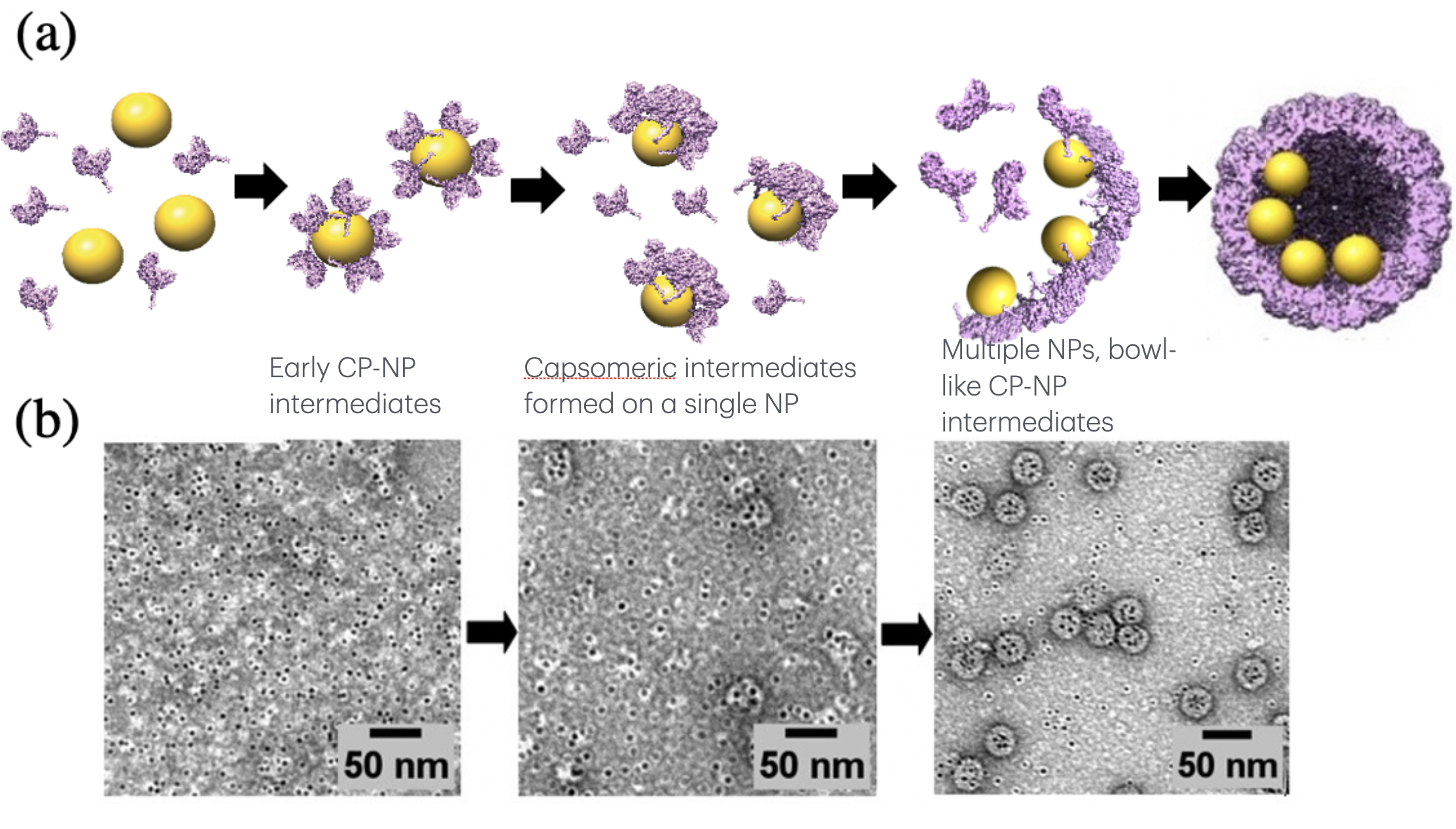}
  \caption{(a) Cartoon of the proposed assembly-line mechanism. (b) Corresponding negative stained micrographs for the assembly stages (with TEGylated NPs).}
  \label{fig:mechanism}
\end{figure}

In summary, the proposed scenario brings to mind an assembly-line mechanism~\cite{Crane1950, morozov2009assembly} where NPs first acquire a number of protein dimers that form a stable intermediate. A new phase follows where multiple CP-NP hetero-oligomers combine to form a cupped NP:CP structure. The VLP is then spontaneously completed by the addition of more heterodimers or even free proteins. Fig.~\ref{fig:mechanism} provides a cartoon of the proposed mechanism for the assembly of VLPs with multiple cores and TEM snapshots of its stages. 

 Similarly to scaffold proteins~\cite{fane2003mechanism,thuman1998role,earnshaw1978structure,salunke1989polymorphism,erickson1981role}, the interacting Au-NP array can rigidify the VLP shell, acting as a backbone, leading to the formation of a capsid consistent in size with the wt BMV, unlike small, flexible oligomers that form strained VLPs with BMV CPs~\cite{bond2020virus} or with CCMV CPs~\cite{maassen2018oligonucleotide}.  Thus, small Au NPs appear to play two key roles: they facilitate assembly nucleation by lowering the electrostatic barrier among components and guide the assembly to completion, ensuring proper virus morphology.

In conclusion, this study advances the understanding of VLP assembly around a sparse set of multiple, rigid, small spherical cargoes, by providing experimental microscopic evidence for intermediates consistent with an ``assembly-line'' mechanism, which can be construed as a sequence of associations of progressively more complex oligomeric structures. Adding a third mechanism to the other two distinct ones previously observed \emph{in vitro} highlights the versatility of the BMV coat protein in dealing with the constraints of the chemical space at assembly. This characteristic further obscures the actual mechanisms by which viruses may assemble \emph{in vivo}, but reveals a potentially broad and potentially useful feature of the capsid proteins of small icosahedral plant viruses. By optimizing parameters such as ionic strength, CP-to-NP ratio, ligand length, and pH, we demonstrated efficient assembly of multicore VLPs having T=3 shell structures as evidenced by atomic force microscopy of single particles. The cryo-EM tomography results indicated a prevalence of NP clustering on one side of the cage at encapsidation, suggesting a role for the NP-NP interaction in guiding morphogenesis. Importantly, for delivery applications, competition experiments pointed to a pronounced NP size selectivity at assembly. These findings may be valuable in the design of future VLP vectors, thus expanding their potential applications in biotechnology and medicine, particularly in synthetic biology and targeted drug delivery.

\clearpage
\section{Materials and Methods}
\subsection{BMV Purification}
Wild-type BMV (wtBMV) was expressed in Nicotiana benthamiana via Agrobacterium-mediated gene delivery according to a previously described protocol~\cite{Gopinath2007}. Seven days after infection, N. benthamiana leaves were homogenized in virus buffer [250 mM NaOAc, 10 mM MgCl$_{2}$, pH 4.5] and then centrifuged at 6000 rpm for 25 minutes using an Eppendorf F-35-6-30 rotor. The supernatant was layered on a 10\% sucrose cushion in virus buffer and centrifuged at 26000 rpm for 3 hours using a Beckman SW 32Ti rotor. The pellets were resuspended in 40\% CsCl (w/v, virus buffer) and were kept overnight in a cold room. After 24 hours, the solution was centrifuged at 10,000 rpm for 10 minutes in the BIO-Rad to remove the undissolved pellets. After that, the supernatant was centrifuged for 24 hours at 45,000 rpm at 4 \textdegree C in a 100 Ti rotor. The resulting virus band was collected and dialyzed against SAMA buffer [50 mM NaOAc, 8 mM Mg(OAc)$_{2}$,  pH 4.6] for 24 hours, with three changes. Virus concentration was measured by UV–Visible spectrophotometer using $\varepsilon$ A (1\%) = 51.5 and the 260/280 absorbance  ratio was $\sim$1.65. 

\subsection{BMV Disassembly} 
Purified virions were dialyzed against disassembly buffer [0.5 M CaCl$_{2}$ (pH 7.4)] for 24-36 hours to precipitate ssRNA. The solution was centrifuged for 30 minutes at 30,000 rpm using a Beckman 70 Ti rotor. The supernatant containing free protein dimers was dialyzed against Tris buffer [10 mM Tris (pH 7.4)] for desalting and then against TKM buffer [1 M KCl, 0.005 MgCl$_{2}$ 0.01 M Tris, pH 7.4] for stabilization. Protein concentration was measured using UV–Visible spectrophotometer  ($\varepsilon$ A (1\%) = 8.20), and the 260/280 absorbance ratio was determined to be $\sim$0.6. The proteins were stored in TKM buffer at 4 \textdegree C. 

\subsection{Synthesis of Gold Nanoparticles (AuNPs), ligand exchange and purification}
AuNPs were synthesized using a previously reported protocol with minor modifications\cite{peng2008}. NPs of average diameter of approximately 3 and 4 nm were synthesized via a burst nucleation method, where the gold precursor was reduced by t-butylamine-borane (TBAB) complex in the presence of the capping agent, olelyamine (OAm). In a typical synthesis, 5 mL of both olelyamine and hexane were stirred with 10mg of HAuCl$_{4}$·3$H_{2}$O under $ N_{2} $ flow for 10 min at 35°C. A reducing solution containing 0.075 mmol of TBAB and 1 mL of hexane and OAm was prepared by sonication and injected into the gold solution. An instantaneous color change was observed from light yellow to reddish brown and the reaction was allowed to proceed for 1 hour. The synthesized NPs were washed with 30 mL of acetone and centrifuged at 7800 rpm for 10 minutes in a F-35-6-30 rotor; this was repeated twice to ensure thorough removal of excess ligands. The precipitates obtained were then redispersed in hexane and characterized by TEM, with a minimum of 300-500 particles counted for each sample to determine the NP size. After synthesis, the NPs were functionalized with thiolated carboxylated tetraethylene glycol (TEG) ligand (HS-C$_{11}$-TEG-CH$_{2}$COOH) or 6-mercaptohexanoic acid ligand at room temperature and centrifuged four times at 45,000 rpm for 40 minutes in a 70 Ti rotor to ensure complete removal of any excess ligands in the solution. The samples were imaged by TEM, and a minimum of 300-500 particles were counted for each sample to measure the size of nanoparticles. NP concentration was obtained from UV-visible absorption measurements performed using a Cary 100 Bio spectrophotometer. First, the total gold concentration was estimated from absorption at 400 nm and then the NP size histogram was used to find the number concentration. 

\subsection{VLP Assembly on functionalized AuNPs}
VLPs were synthesized by using a nanoparticle templated self-assembly method as reported previously~\cite{Vieweger2018} with modifications in protein to nanoparticle ratio as indicated in results section.  The volumes of freshly prepared proteins (within 1 or 2 days after the virus dissociation process) were diluted in TKM buffer to ~0.5 mg/mL and the functionalized NPs were added to the solution. The sample was dialyzed against virus reassembly buffer (0.05 M Tris, 0.05 M NaCl, 0.01 M KCl, 0.005 M MgCl$_{2}$, pH 7.2) for at least 24 h at 4 \textdegree C followed by dialysis in SAMA buffer (0.05 M NaOAc, 0.008 M Mg(OAc)$_{2}$, pH 4.6) for an additional 24 h at 4 \textdegree C.

\subsection{Dynamic Light Scattering and Zetapotential Measurements}
Dynamic light scattering (DLS) and Zetapotential measurements were carried out with a Zetasizer Nano-S instrument (Malvern). DLS was performed to measure the hydrodynamic diameter of suspended particles in a carrier solution. Typically, the diluted sample was sonicated for about 10 to 20 min followed by filtration with a 0.2 mm syringe filter prior to measurement. The samples were loaded into low volume disposable sizing cuvettes. The experiment was carried out at 25\textdegree C. The measurement duration was set to be determined automatically, and data were averaged from at least three runs. Intensity and volume distributions of the particle sizes were recorded and analyzed. The sample was diluted 1000$\times$ in potassium nitrate and the zeta potential was measured in a folded capillary cell. Data was processed using the absorption of bulk gold, the indices of refraction of gold and potassium nitrate, and the viscosity of potassium nitrate. The Smoluchowski approximation was used to convert the electrophoretic mobility to a zeta potential.

\subsection{Gel Electrophoresis}
Gel electrophoresis was used to provide evidence of anionic functional group attachment to the NP surface. The electrophoretic matrix was made of agarose gel (2\%) in a TBE buffer [Tris base (0.4 M), boric acid (0.45 M), EDTA (10 mM)]. The sample was run at a constant voltage of 7.5 V/cm, and the mobility of the sample was assessed by monitoring the movement of the bands over time.

\subsection{Transmission Electron Microscopy}
Electron-transparent samples were prepared by placing 10$\mu$L of a dilute sample on a carbon-coated copper grid. After 10 min, the excess solution on the grid was removed with filter paper. The sample was stained with 10$\mu$L of 2\% uranyl acetate for 10 min, and the excess solution was removed by blotting with filter paper. The sample was then left to dry for several minutes. The images were acquired at an accelerating voltage of 80 kV on a JEOL JEM1010 and JEM1400plus transmission electron microscope and analyzed with the ImageJ Processing Toolkit for overall morphological characterization and to estimate particle diameters and the number of encapsulated gold NPs. 

\subsection{Cryo-EM}\label{Cryo-EM}
Cryo-EM samples were prepared by applying 5 $\mu$L of the sample solution onto a glow-discharged holey carbon film-coated copper grid (Quantifoil R1.2/1.3). The grid was then frozen using an FEI Vitrobot (Mark IV) under the following conditions: temperature of 22 \textdegree C, application of +2 force, a 1 second wait time, 4 seconds of blotting time, and 100\% humidity. The frozen hydrated cryo-EM grids were then transferred to the 200-kV Talos Arctica microscope (Thermo Fisher Scientific) for the analysis of virus-like particles (VLPs). Images were acquired in electron counting mode using the Falcon 3 camera on the Talos Arctica, at a magnification of 150,000$\times$. A total of forty frames were captured and aligned using EPU software. Each image was exposed to a total dose of approximately 35 e$^{-}$/Å$^{2}$, with a frame dose rate of 0.875 e$^{-}$/Å$^2$. The effective pixel size for the acquired data was 0.9693 Å.

\subsection{Tomography}
The sample was frozen according to the Cryo-EM protocol, with the addition of 11.6 nm AuNP fiducial markers in the solution. Cryo-ET imaging was performed using a Thermo-Fisher Talos Arctica microscope at 200 kV, equipped with a TFS Falcon 3 camera. In cryo-ET, the specimen is tilted during acquisition, capturing 2D projections from various angles, which are then reconstructed into a 3D tomogram. Tilt series (TS) were acquired using serialEM and the Falcon 3 camera in linear mode. Images were recorded at a nominal magnification of 92,000$\times$ magnification with a -6 $\mu$m defocus, corresponding to pixel size of 1.65 Å. The tilt angles ranged from 60$^{\circ}$ to -60$^{\circ}$ with 3$^{\circ}$ increments. Each tilt series image was exposed to a total dose of approximately 105 e$^{-}$/Å$^{2}$, with a frame dose rate of 2.15 e$^{-}$/Å$^{2}$. The tilt series were aligned and reconstructed using the IMOD software package\cite{Mastronarde2017AutomatedTS}, using the gold fiducial markers to align the TS images.

\begin{suppinfo}

\end{suppinfo}
Supporting Information is available from the Online Library or from the author. TEM images of T=1 VLPs formed with 4.7 nm AuNPs, AFM images of partial capsid fragments adsorbed on HOPG, TEM images of VLPs assembled with HA-functionalized AuNPs, Cryo-EM images of VLPs assembled around HA and TEG AuNPs, surface charge comparison of HA and TEG functionalized AuNPs (PDF). Tomographic reconstructions of VLPs containing clustered NP cargo (Supporting Videos S1–S3).

\begin{acknowledgement}
The work was partly supported by the U.S. Army Research Office through award \#W911NF1310490. R.Z. acknowledge support from NSF DMR-2131963 and the University of California Multicampus Research Programs and Initiatives (Grant No. M21PR3267). The authors gratefully acknowledge Indiana University at Bloomington for access to the Electron Microscopy Center SCR\_017845 and the Nanoscience Characterization Facility.
\end{acknowledgement}

\bibliography{bibliography}

\providecommand{\latin}[1]{#1}
\makeatletter
\providecommand{\doi}
  {\begingroup\let\do\@makeother\dospecials
  \catcode`\{=1 \catcode`\}=2 \doi@aux}
\providecommand{\doi@aux}[1]{\endgroup\texttt{#1}}
\makeatother
\providecommand*\mcitethebibliography{\thebibliography}
\csname @ifundefined\endcsname{endmcitethebibliography}
  {\let\endmcitethebibliography\endthebibliography}{}
\begin{mcitethebibliography}{111}
\providecommand*\natexlab[1]{#1}
\providecommand*\mciteSetBstSublistMode[1]{}
\providecommand*\mciteSetBstMaxWidthForm[2]{}
\providecommand*\mciteBstWouldAddEndPuncttrue
  {\def\EndOfBibitem{\unskip.}}
\providecommand*\mciteBstWouldAddEndPunctfalse
  {\let\EndOfBibitem\relax}
\providecommand*\mciteSetBstMidEndSepPunct[3]{}
\providecommand*\mciteSetBstSublistLabelBeginEnd[3]{}
\providecommand*\EndOfBibitem{}
\mciteSetBstSublistMode{f}
\mciteSetBstMaxWidthForm{subitem}{(\alph{mcitesubitemcount})}
\mciteSetBstSublistLabelBeginEnd
  {\mcitemaxwidthsubitemform\space}
  {\relax}
  {\relax}

\bibitem[Crane(1950)]{Crane1950}
Crane,~H.~R. Principles and Problems of Biological Growth. \emph{Sci. Mon.}
  \textbf{1950}, \emph{70}, 376--389\relax
\mciteBstWouldAddEndPuncttrue
\mciteSetBstMidEndSepPunct{\mcitedefaultmidpunct}
{\mcitedefaultendpunct}{\mcitedefaultseppunct}\relax
\EndOfBibitem
\bibitem[Watson and Crick(1953)Watson, and Crick]{WatsonCrick1953}
Watson,~J.~D.; Crick,~F.~H. Molecular Structure of Nucleic Acids: A Structure
  for Deoxyribose Nucleic Acid. \emph{Nature} \textbf{1953}, \emph{171},
  737--738\relax
\mciteBstWouldAddEndPuncttrue
\mciteSetBstMidEndSepPunct{\mcitedefaultmidpunct}
{\mcitedefaultendpunct}{\mcitedefaultseppunct}\relax
\EndOfBibitem
\bibitem[Franklin and Gosling(1953)Franklin, and Gosling]{FranklinGosling1953}
Franklin,~R.~E.; Gosling,~R.~G. Molecular Configuration in Sodium
  Thymonucleate. \emph{Nature} \textbf{1953}, \emph{171}, 740--741\relax
\mciteBstWouldAddEndPuncttrue
\mciteSetBstMidEndSepPunct{\mcitedefaultmidpunct}
{\mcitedefaultendpunct}{\mcitedefaultseppunct}\relax
\EndOfBibitem
\bibitem[Wilkins \latin{et~al.}(1953)Wilkins, Stokes, and Wilson]{Wilkins1953}
Wilkins,~M. H.~F.; Stokes,~A.~R.; Wilson,~H.~R. Molecular Structure of
  Deoxypentose Nucleic Acids. \emph{Nature} \textbf{1953}, \emph{171},
  738--740\relax
\mciteBstWouldAddEndPuncttrue
\mciteSetBstMidEndSepPunct{\mcitedefaultmidpunct}
{\mcitedefaultendpunct}{\mcitedefaultseppunct}\relax
\EndOfBibitem
\bibitem[Oosawa and Asakura(1975)Oosawa, and Asakura]{oosawa1975thermodynamics}
Oosawa,~F.; Asakura,~S. \emph{Thermodynamics of the Polymerization of Protein};
  Academic Press, 1975; Vol.~20\relax
\mciteBstWouldAddEndPuncttrue
\mciteSetBstMidEndSepPunct{\mcitedefaultmidpunct}
{\mcitedefaultendpunct}{\mcitedefaultseppunct}\relax
\EndOfBibitem
\bibitem[Zlotnick(2003)]{Zlotnick2003}
Zlotnick,~A. {Are Weak Protein-Protein Interactions the General Rule in Capsid
  Assembly?} \emph{Virology} \textbf{2003}, \emph{315}, 269--274\relax
\mciteBstWouldAddEndPuncttrue
\mciteSetBstMidEndSepPunct{\mcitedefaultmidpunct}
{\mcitedefaultendpunct}{\mcitedefaultseppunct}\relax
\EndOfBibitem
\bibitem[Curk \latin{et~al.}(2017)Curk, Dobnikar, and Frenkel]{Curk2017}
Curk,~T.; Dobnikar,~J.; Frenkel,~D. In \emph{Multivalency - Concepts, Research,
  Applications}; Huskens,~J., Prins,~L.~J., Haag,~R., Ravoo,~B.~J., Eds.;
  Wiley, 2017; Chapter 3, pp 75--101\relax
\mciteBstWouldAddEndPuncttrue
\mciteSetBstMidEndSepPunct{\mcitedefaultmidpunct}
{\mcitedefaultendpunct}{\mcitedefaultseppunct}\relax
\EndOfBibitem
\bibitem[Zandi \latin{et~al.}(2006)Zandi, van~der Schoot, Reguera, Kegel, and
  Reiss]{zandi2006classical}
Zandi,~R.; van~der Schoot,~P.; Reguera,~D.; Kegel,~W.; Reiss,~H. Classical
  Nucleation Theory of Virus Capsids. \emph{Biophys. J.} \textbf{2006},
  \emph{90}, 1939--1948\relax
\mciteBstWouldAddEndPuncttrue
\mciteSetBstMidEndSepPunct{\mcitedefaultmidpunct}
{\mcitedefaultendpunct}{\mcitedefaultseppunct}\relax
\EndOfBibitem
\bibitem[Chung \latin{et~al.}(2020)Chung, Cai, and Steinmetz]{Chung2020}
Chung,~Y.~H.; Cai,~H.; Steinmetz,~N.~F. Viral Nanoparticles for Drug Delivery,
  Imaging, Immunotherapy, and Theranostic Applications. \emph{Adv. Drug
  Delivery Rev.} \textbf{2020}, \emph{156}, 214--235\relax
\mciteBstWouldAddEndPuncttrue
\mciteSetBstMidEndSepPunct{\mcitedefaultmidpunct}
{\mcitedefaultendpunct}{\mcitedefaultseppunct}\relax
\EndOfBibitem
\bibitem[Tsvetkova and Dragnea(2015)Tsvetkova, and Dragnea]{Tsvetkova2015}
Tsvetkova,~I.~B.; Dragnea,~B.~G. In \emph{Protein Cages: Methods and
  Protocols}; Orner,~B., Ed.; Methods Mol. Bio.; Springer, 2015; Vol. 1252; pp
  1--15\relax
\mciteBstWouldAddEndPuncttrue
\mciteSetBstMidEndSepPunct{\mcitedefaultmidpunct}
{\mcitedefaultendpunct}{\mcitedefaultseppunct}\relax
\EndOfBibitem
\bibitem[Saxena \latin{et~al.}(2016)Saxena, He, Malyutin, Datta, Rein, Bond,
  Jarrold, Spilotros, Svergun, Douglas, and Dragnea]{Saxena2016}
Saxena,~P.; He,~L.; Malyutin,~A.; Datta,~S. A.~K.; Rein,~A.; Bond,~K.~M.;
  Jarrold,~M.~F.; Spilotros,~A.; Svergun,~D.; Douglas,~T.; Dragnea,~B. Virus
  Matryoshka: A Bacteriophage Particle-Guided Molecular Assembly Approach to a
  Monodisperse Model of the Immature Human Immunodeficiency Virus. \emph{Small}
  \textbf{2016}, \emph{12}, 5862--5872\relax
\mciteBstWouldAddEndPuncttrue
\mciteSetBstMidEndSepPunct{\mcitedefaultmidpunct}
{\mcitedefaultendpunct}{\mcitedefaultseppunct}\relax
\EndOfBibitem
\bibitem[Xie \latin{et~al.}(2021)Xie, Tsvetkova, Liu, Ye, Hewavitharanage,
  Dragnea, and Cadena-Nava]{Xie2021}
Xie,~A.; Tsvetkova,~I.; Liu,~Y.; Ye,~X.; Hewavitharanage,~P.; Dragnea,~B.;
  Cadena-Nava,~R.~D. Hydrophobic Cargo Encapsulation into Virus Protein Cages
  by Self-Assembly in an Aprotic Organic Solvent. \emph{Bioconjug. Chem.}
  \textbf{2021}, \emph{32}, 2366--2376\relax
\mciteBstWouldAddEndPuncttrue
\mciteSetBstMidEndSepPunct{\mcitedefaultmidpunct}
{\mcitedefaultendpunct}{\mcitedefaultseppunct}\relax
\EndOfBibitem
\bibitem[Uchida \latin{et~al.}(2022)Uchida, Manzo, Echeveria, Jimenez, and
  Lovell]{Uchida2022}
Uchida,~M.; Manzo,~E.; Echeveria,~D.; Jimenez,~S.; Lovell,~L. Harnessing
  Physicochemical Properties of Virus Capsids for Designing Enzyme Confined
  Nanocompartments. \emph{Curr. Opin. Virol.} \textbf{2022}, \emph{52},
  250--257\relax
\mciteBstWouldAddEndPuncttrue
\mciteSetBstMidEndSepPunct{\mcitedefaultmidpunct}
{\mcitedefaultendpunct}{\mcitedefaultseppunct}\relax
\EndOfBibitem
\bibitem[Seitz \latin{et~al.}(2023)Seitz, Saarinen, Kumpula, McNeale,
  Anaya-Plaza, Lampinen, Hytoenen, Sainsbury, Cornelissen, Linko, Huiskonen,
  and Kostiainen]{Seitz2023}
Seitz,~I.; Saarinen,~S.; Kumpula,~E.-P.; McNeale,~D.; Anaya-Plaza,~E.;
  Lampinen,~V.; Hytoenen,~V.~P.; Sainsbury,~F.; Cornelissen,~J. J. L.~M.;
  Linko,~V.; Huiskonen,~J.~T.; Kostiainen,~M.~A. DNA-Origami-Directed Virus
  Capsid Polymorphism. \emph{Nat. Nanotechnol.} \textbf{2023}, \emph{18},
  1205+\relax
\mciteBstWouldAddEndPuncttrue
\mciteSetBstMidEndSepPunct{\mcitedefaultmidpunct}
{\mcitedefaultendpunct}{\mcitedefaultseppunct}\relax
\EndOfBibitem
\bibitem[Omole \latin{et~al.}(2024)Omole, Zhao, Chang-Liao, de~Oliveira, Boone,
  Sutorus, Sack, Varner, Fiering, and Steinmetz]{Boone2024}
Omole,~A.~O.; Zhao,~Z.; Chang-Liao,~S.; de~Oliveira,~J. F.~A.; Boone,~C.~E.;
  Sutorus,~L.; Sack,~M.; Varner,~J.; Fiering,~S.~N.; Steinmetz,~N.~F. Virus
  Nanotechnology for Intratumoural Immunotherapy. \emph{Nat. Rev. Bioeng.}
  \textbf{2024}, \emph{2}, 916--929\relax
\mciteBstWouldAddEndPuncttrue
\mciteSetBstMidEndSepPunct{\mcitedefaultmidpunct}
{\mcitedefaultendpunct}{\mcitedefaultseppunct}\relax
\EndOfBibitem
\bibitem[Zlotnick and Mukhopadhyay(2011)Zlotnick, and
  Mukhopadhyay]{Zlotnick2011}
Zlotnick,~A.; Mukhopadhyay,~S. {Virus assembly, allostery and antivirals.}
  \emph{Trends Microbiol.} \textbf{2011}, \emph{19}, 14--23\relax
\mciteBstWouldAddEndPuncttrue
\mciteSetBstMidEndSepPunct{\mcitedefaultmidpunct}
{\mcitedefaultendpunct}{\mcitedefaultseppunct}\relax
\EndOfBibitem
\bibitem[Bond \latin{et~al.}(2020)Bond, Tsvetkova, Wang, Jarrold, and
  Dragnea]{bond2020virus}
Bond,~K.; Tsvetkova,~I.~B.; Wang,~J. C.-Y.; Jarrold,~M.~F.; Dragnea,~B. Virus
  Assembly Pathways: Straying Away but Not Too Far. \emph{Small} \textbf{2020},
  \emph{16}, 2004475\relax
\mciteBstWouldAddEndPuncttrue
\mciteSetBstMidEndSepPunct{\mcitedefaultmidpunct}
{\mcitedefaultendpunct}{\mcitedefaultseppunct}\relax
\EndOfBibitem
\bibitem[Toropova \latin{et~al.}(2008)Toropova, Basnak, Twarock, Stockley, and
  Ranson]{Toropova2008}
Toropova,~K.; Basnak,~G.; Twarock,~R.; Stockley,~P.~G.; Ranson,~N.~A. {The
  Three-dimensional Structure of Genomic RNA in Bacteriophage MS2: Implications
  for Assembly}. \emph{J. Mol. Biol.} \textbf{2008}, \emph{375}, 824--836\relax
\mciteBstWouldAddEndPuncttrue
\mciteSetBstMidEndSepPunct{\mcitedefaultmidpunct}
{\mcitedefaultendpunct}{\mcitedefaultseppunct}\relax
\EndOfBibitem
\bibitem[Stockley \latin{et~al.}(2013)Stockley, Ranson, and
  Twarock]{Stockley2013}
Stockley,~P.~G.; Ranson,~N.~A.; Twarock,~R. A New Paradigm for the Roles of the
  Genome in ssRNA Viruses. \emph{Future Virol.} \textbf{2013}, \emph{8},
  531--543\relax
\mciteBstWouldAddEndPuncttrue
\mciteSetBstMidEndSepPunct{\mcitedefaultmidpunct}
{\mcitedefaultendpunct}{\mcitedefaultseppunct}\relax
\EndOfBibitem
\bibitem[McDonald and Patton(2011)McDonald, and Patton]{McDonald2011}
McDonald,~S.~M.; Patton,~J.~T. Assortment and Packaging of the Segmented
  Rotavirus genome. \emph{Trends Microbiol.} \textbf{2011}, \emph{19},
  136--144\relax
\mciteBstWouldAddEndPuncttrue
\mciteSetBstMidEndSepPunct{\mcitedefaultmidpunct}
{\mcitedefaultendpunct}{\mcitedefaultseppunct}\relax
\EndOfBibitem
\bibitem[Noueiry and Ahlquist(2003)Noueiry, and Ahlquist]{noueiry2003brome}
Noueiry,~A.~O.; Ahlquist,~P. Brome Mosaic Virus RNA Replication: Revealing the
  Role of the Host in RNA Virus Replication. \emph{Annu. Rev. Phytopathol.}
  \textbf{2003}, \emph{41}, 77--98\relax
\mciteBstWouldAddEndPuncttrue
\mciteSetBstMidEndSepPunct{\mcitedefaultmidpunct}
{\mcitedefaultendpunct}{\mcitedefaultseppunct}\relax
\EndOfBibitem
\bibitem[Schwartz \latin{et~al.}(2002)Schwartz, Chen, Janda, Sullivan, den
  Boon, and Ahlquist]{schwartz2002positive}
Schwartz,~M.; Chen,~J.; Janda,~M.; Sullivan,~M.; den Boon,~J.; Ahlquist,~P. A
  Positive-Strand Rna Virus Replication Complex Parallels Form and Function of
  Retrovirus Capsids. \emph{Mol. Cell} \textbf{2002}, \emph{9}, 505--514\relax
\mciteBstWouldAddEndPuncttrue
\mciteSetBstMidEndSepPunct{\mcitedefaultmidpunct}
{\mcitedefaultendpunct}{\mcitedefaultseppunct}\relax
\EndOfBibitem
\bibitem[Schwartz \latin{et~al.}(2004)Schwartz, Chen, Lee, Janda, and
  Ahlquist]{schwartz2004alternate}
Schwartz,~M.; Chen,~J.; Lee,~W.-M.; Janda,~M.; Ahlquist,~P. Alternate,
  Virus-Induced Membrane Rearrangements Support Positive-Strand RNA Virus
  Genome Replication. \emph{Proc. Nat. Acad. Sci.} \textbf{2004}, \emph{101},
  11263--11268\relax
\mciteBstWouldAddEndPuncttrue
\mciteSetBstMidEndSepPunct{\mcitedefaultmidpunct}
{\mcitedefaultendpunct}{\mcitedefaultseppunct}\relax
\EndOfBibitem
\bibitem[Miller \latin{et~al.}(1985)Miller, Dreher, and
  Hall]{miller1985synthesis}
Miller,~W.; Dreher,~T.; Hall,~T. Synthesis of Brome Mosaic Virus Subgenomic RNA
  In Vitro by Internal Initiation on (--)-Sense Genomic RNA. \emph{Nature}
  \textbf{1985}, \emph{313}, 68--70\relax
\mciteBstWouldAddEndPuncttrue
\mciteSetBstMidEndSepPunct{\mcitedefaultmidpunct}
{\mcitedefaultendpunct}{\mcitedefaultseppunct}\relax
\EndOfBibitem
\bibitem[W{\"{o}}rsd{\"{o}}rfer \latin{et~al.}(2011)W{\"{o}}rsd{\"{o}}rfer,
  Woycechowsky, and Hilvert]{Worsdorfer2011}
W{\"{o}}rsd{\"{o}}rfer,~B.; Woycechowsky,~K.~J.; Hilvert,~D. Directed Evolution
  of a Protein Container. \emph{Science} \textbf{2011}, \emph{331},
  589--592\relax
\mciteBstWouldAddEndPuncttrue
\mciteSetBstMidEndSepPunct{\mcitedefaultmidpunct}
{\mcitedefaultendpunct}{\mcitedefaultseppunct}\relax
\EndOfBibitem
\bibitem[Tetter \latin{et~al.}(2021)Tetter, Terasaka, Steinauer, Bingham,
  Clark, Scott, Patel, Leibundgut, Wroblewski, Ban, Stockley, Twarock, and
  Hilvert]{Tetter2021}
Tetter,~S.; Terasaka,~N.; Steinauer,~A.; Bingham,~R.~J.; Clark,~S.; Scott,~A.
  J.~P.; Patel,~N.; Leibundgut,~M.; Wroblewski,~E.; Ban,~N.; Stockley,~P.~G.;
  Twarock,~R.; Hilvert,~D. Evolution of a Virus-like Architecture and Packaging
  Mechanism in a Repurposed Bacterial Protein. \emph{Science} \textbf{2021},
  \emph{372}, 1220--1224\relax
\mciteBstWouldAddEndPuncttrue
\mciteSetBstMidEndSepPunct{\mcitedefaultmidpunct}
{\mcitedefaultendpunct}{\mcitedefaultseppunct}\relax
\EndOfBibitem
\bibitem[Patterson \latin{et~al.}(2014)Patterson, Schwarz, Waters, Gedeon, and
  Douglas]{Patterson2014}
Patterson,~D.~P.; Schwarz,~B.; Waters,~R.~S.; Gedeon,~T.; Douglas,~T.
  Encapsulation of an Enzyme Cascade within the Bacteriophage P22 Virus-Like
  Particle. \emph{ACS Chem. Biol.} \textbf{2014}, \emph{9}, 359--365\relax
\mciteBstWouldAddEndPuncttrue
\mciteSetBstMidEndSepPunct{\mcitedefaultmidpunct}
{\mcitedefaultendpunct}{\mcitedefaultseppunct}\relax
\EndOfBibitem
\bibitem[Guerrero \latin{et~al.}(2015)Guerrero, Bahmani, Singh, Vullev, Kundra,
  and Anvari]{Guerrero2015}
Guerrero,~Y.~A.; Bahmani,~B.; Singh,~S.~P.; Vullev,~V.~I.; Kundra,~V.;
  Anvari,~B. Virus-Resembling Nano-Structures for Near Infrared Fluorescence
  Imaging of Ovarian Cancer HER2 Receptors. \emph{Nanotechnology}
  \textbf{2015}, \emph{26}\relax
\mciteBstWouldAddEndPuncttrue
\mciteSetBstMidEndSepPunct{\mcitedefaultmidpunct}
{\mcitedefaultendpunct}{\mcitedefaultseppunct}\relax
\EndOfBibitem
\bibitem[Le \latin{et~al.}(2017)Le, Lee, Shukla, Commandeur, and
  Steinmetz]{Le2017}
Le,~D. H.~T.; Lee,~K.~L.; Shukla,~S.; Commandeur,~U.; Steinmetz,~N.~F. Potato
  virus X, a filamentous plant viral nanoparticle for doxorubicin delivery in
  cancer therapy. \emph{Nanoscale} \textbf{2017}, \emph{9}, 2348--2357\relax
\mciteBstWouldAddEndPuncttrue
\mciteSetBstMidEndSepPunct{\mcitedefaultmidpunct}
{\mcitedefaultendpunct}{\mcitedefaultseppunct}\relax
\EndOfBibitem
\bibitem[Chen \latin{et~al.}(2006)Chen, Daniel, Quinkert, De, Stein, Bowman,
  Chipman, Rotello, Kao, and Dragnea]{chen2006nanoparticle}
Chen,~C.; Daniel,~M.-C.; Quinkert,~Z.~T.; De,~M.; Stein,~B.; Bowman,~V.~D.;
  Chipman,~P.~R.; Rotello,~V.~M.; Kao,~C.~C.; Dragnea,~B.
  Nanoparticle-Templated Assembly of Viral Protein Cages. \emph{Nano Letters.}
  \textbf{2006}, \emph{6}, 611--615\relax
\mciteBstWouldAddEndPuncttrue
\mciteSetBstMidEndSepPunct{\mcitedefaultmidpunct}
{\mcitedefaultendpunct}{\mcitedefaultseppunct}\relax
\EndOfBibitem
\bibitem[Benjamin \latin{et~al.}(2018)Benjamin, Chen, Kang, Wilson, Li,
  Nielsen, Qin, and Gassensmith]{Benjamin2018}
Benjamin,~C.~E.; Chen,~Z.; Kang,~P.; Wilson,~B.~A.; Li,~N.; Nielsen,~S.~O.;
  Qin,~Z.; Gassensmith,~J.~J. Site-Selective Nucleation and Size Control of
  Gold Nanoparticle Photothermal Antennae on the Pore Structures of a Virus.
  \emph{JOURNAL OF THE AMERICAN CHEMICAL SOCIETY} \textbf{2018}, \emph{140},
  17226--17233\relax
\mciteBstWouldAddEndPuncttrue
\mciteSetBstMidEndSepPunct{\mcitedefaultmidpunct}
{\mcitedefaultendpunct}{\mcitedefaultseppunct}\relax
\EndOfBibitem
\bibitem[McNeale \latin{et~al.}(2023)McNeale, Esquirol, Okada, Strampel,
  Dashti, Rehm, Douglas, Vickers, and Sainsbury]{mcneale2023tunable}
McNeale,~D.; Esquirol,~L.; Okada,~S.; Strampel,~S.; Dashti,~N.; Rehm,~B.;
  Douglas,~T.; Vickers,~C.; Sainsbury,~F. Tunable In Vivo Colocalization of
  Enzymes within P22 Capsid-Based Nanoreactors. \emph{ACS Applied Materials \&
  Interfaces} \textbf{2023}, \emph{15}, 17705--17715\relax
\mciteBstWouldAddEndPuncttrue
\mciteSetBstMidEndSepPunct{\mcitedefaultmidpunct}
{\mcitedefaultendpunct}{\mcitedefaultseppunct}\relax
\EndOfBibitem
\bibitem[Sharma \latin{et~al.}(2017)Sharma, Uchida, Miettinen, and
  Douglas]{sharma2017modular}
Sharma,~J.; Uchida,~M.; Miettinen,~H.~M.; Douglas,~T. Modular interior loading
  and exterior decoration of a virus-like particle. \emph{Nanoscale}
  \textbf{2017}, \emph{9}, 10420--10430\relax
\mciteBstWouldAddEndPuncttrue
\mciteSetBstMidEndSepPunct{\mcitedefaultmidpunct}
{\mcitedefaultendpunct}{\mcitedefaultseppunct}\relax
\EndOfBibitem
\bibitem[Oerlemans \latin{et~al.}(2021)Oerlemans, Timmermans, and van
  Hest]{Oerlemans2021}
Oerlemans,~R. A. J.~F.; Timmermans,~S. B. P.~E.; van Hest,~J. C.~M. Artificial
  Organelles: Towards Adding or Restoring Intracellular Activity.
  \emph{ChemBioChem} \textbf{2021}, \emph{22}, 2051--2078\relax
\mciteBstWouldAddEndPuncttrue
\mciteSetBstMidEndSepPunct{\mcitedefaultmidpunct}
{\mcitedefaultendpunct}{\mcitedefaultseppunct}\relax
\EndOfBibitem
\bibitem[Kong \latin{et~al.}(2015)Kong, Liu, Jia, Wu, Wu, Chen, and
  Fang]{kong2015pokemon}
Kong,~J.; Liu,~X.; Jia,~J.; Wu,~J.; Wu,~N.; Chen,~J.; Fang,~F. Pokemon siRNA
  delivery mediated by RGD-modified HBV core protein suppressed the growth of
  hepatocellular carcinoma. \emph{Human Gene Therapy Methods} \textbf{2015},
  \emph{26}, 175--180\relax
\mciteBstWouldAddEndPuncttrue
\mciteSetBstMidEndSepPunct{\mcitedefaultmidpunct}
{\mcitedefaultendpunct}{\mcitedefaultseppunct}\relax
\EndOfBibitem
\bibitem[Parent \latin{et~al.}(2005)Parent, Doyle, Anderson, and
  Teschke]{parent2005electrostatic}
Parent,~K.~N.; Doyle,~S.~M.; Anderson,~E.; Teschke,~C.~M. Electrostatic
  interactions govern both nucleation and elongation during phage P22 procapsid
  assembly. \emph{Virology} \textbf{2005}, \emph{340}, 33--45\relax
\mciteBstWouldAddEndPuncttrue
\mciteSetBstMidEndSepPunct{\mcitedefaultmidpunct}
{\mcitedefaultendpunct}{\mcitedefaultseppunct}\relax
\EndOfBibitem
\bibitem[Parent \latin{et~al.}(2006)Parent, Zlotnick, and
  Teschke]{parent2006quantitative}
Parent,~K.~N.; Zlotnick,~A.; Teschke,~C.~M. Quantitative Analysis of
  Multi-Component Spherical Virus Assembly: Scaffolding Protein Contributes to
  the Global Stability of Phage P22 Procapsids. \emph{J. of Mol. Bio.}
  \textbf{2006}, \emph{359}, 1097--1106\relax
\mciteBstWouldAddEndPuncttrue
\mciteSetBstMidEndSepPunct{\mcitedefaultmidpunct}
{\mcitedefaultendpunct}{\mcitedefaultseppunct}\relax
\EndOfBibitem
\bibitem[Teschke and Parent(2010)Teschke, and Parent]{teschke2010let}
Teschke,~C.~M.; Parent,~K.~N. ‘Let the phage do the work’: using the phage
  P22 coat protein structures as a framework to understand its folding and
  assembly mutants. \emph{Virology} \textbf{2010}, \emph{401}, 119--130\relax
\mciteBstWouldAddEndPuncttrue
\mciteSetBstMidEndSepPunct{\mcitedefaultmidpunct}
{\mcitedefaultendpunct}{\mcitedefaultseppunct}\relax
\EndOfBibitem
\bibitem[Fane and Prevelige~Jr(2003)Fane, and Prevelige~Jr]{fane2003mechanism}
Fane,~B.~A.; Prevelige~Jr,~P.~E. Mechanism of Scaffolding-Assisted Viral
  Assembly. \emph{Adv. Protein Chem.} \textbf{2003}, \emph{64}, 259--299\relax
\mciteBstWouldAddEndPuncttrue
\mciteSetBstMidEndSepPunct{\mcitedefaultmidpunct}
{\mcitedefaultendpunct}{\mcitedefaultseppunct}\relax
\EndOfBibitem
\bibitem[Panahandeh \latin{et~al.}(2022)Panahandeh, Li, Dragnea, and
  Zandi]{panahandeh2022virus}
Panahandeh,~S.; Li,~S.; Dragnea,~B.; Zandi,~R. Virus Assembly Pathways Inside a
  Host Cell. \emph{ACS Nano} \textbf{2022}, \emph{16}, 317--327\relax
\mciteBstWouldAddEndPuncttrue
\mciteSetBstMidEndSepPunct{\mcitedefaultmidpunct}
{\mcitedefaultendpunct}{\mcitedefaultseppunct}\relax
\EndOfBibitem
\bibitem[Prevelige \latin{et~al.}(1993)Prevelige, Thomas, and
  King]{prevelige1993nucleation}
Prevelige,~P.; Thomas,~D.; King,~J. Nucleation and Growth Phases in the
  Polymerization of Coat and Scaffolding Subunits into Icosahedral Procapsid
  Shells. \emph{Biophys. J.} \textbf{1993}, \emph{64}, 824--835\relax
\mciteBstWouldAddEndPuncttrue
\mciteSetBstMidEndSepPunct{\mcitedefaultmidpunct}
{\mcitedefaultendpunct}{\mcitedefaultseppunct}\relax
\EndOfBibitem
\bibitem[Zlotnick(1994)]{zlotnick1994build}
Zlotnick,~A. To Build a Virus Capsid: An Equilibrium Model of the Self Assembly
  of Polyhedral Protein Complexes. \emph{J. Mol. Biol.} \textbf{1994},
  \emph{241}, 59--67\relax
\mciteBstWouldAddEndPuncttrue
\mciteSetBstMidEndSepPunct{\mcitedefaultmidpunct}
{\mcitedefaultendpunct}{\mcitedefaultseppunct}\relax
\EndOfBibitem
\bibitem[Kler \latin{et~al.}(2012)Kler, Asor, Li, Ginsburg, Harries, Oppenheim,
  Zlotnick, and Raviv]{kler2012rna}
Kler,~S.; Asor,~R.; Li,~C.; Ginsburg,~A.; Harries,~D.; Oppenheim,~A.;
  Zlotnick,~A.; Raviv,~U. RNA Encapsidation by SV40-Derived Nanoparticles
  Follows a Rapid Two-State Mechanism. \emph{J. Am. Chem. Soc.} \textbf{2012},
  \emph{134}, 8823--8830\relax
\mciteBstWouldAddEndPuncttrue
\mciteSetBstMidEndSepPunct{\mcitedefaultmidpunct}
{\mcitedefaultendpunct}{\mcitedefaultseppunct}\relax
\EndOfBibitem
\bibitem[Hagan(2009)]{hagan2009theory}
Hagan,~M.~F. A Theory for Viral Capsid Assembly around Electrostatic Cores.
  \emph{J. Chem. Phys.} \textbf{2009}, \emph{130}\relax
\mciteBstWouldAddEndPuncttrue
\mciteSetBstMidEndSepPunct{\mcitedefaultmidpunct}
{\mcitedefaultendpunct}{\mcitedefaultseppunct}\relax
\EndOfBibitem
\bibitem[Kler \latin{et~al.}(2013)Kler, Wang, Dhason, Oppenheim, and
  Zlotnick]{kler2013scaffold}
Kler,~S.; Wang,~J. C.-Y.; Dhason,~M.; Oppenheim,~A.; Zlotnick,~A. Scaffold
  Properties are a Key Determinant of the Size and Shape of Self-Assembled
  Virus-Derived Particles. \emph{ACS Chem. Biol.} \textbf{2013}, \emph{8},
  2753--2761\relax
\mciteBstWouldAddEndPuncttrue
\mciteSetBstMidEndSepPunct{\mcitedefaultmidpunct}
{\mcitedefaultendpunct}{\mcitedefaultseppunct}\relax
\EndOfBibitem
\bibitem[Zlotnick \latin{et~al.}(2013)Zlotnick, Porterfield, and
  Wang]{zlotnick2013build}
Zlotnick,~A.; Porterfield,~J.~Z.; Wang,~J. C.-Y. To Build a Virus on a Nucleic
  Acid Substrate. \emph{Biophys. J.} \textbf{2013}, \emph{104},
  1595--1604\relax
\mciteBstWouldAddEndPuncttrue
\mciteSetBstMidEndSepPunct{\mcitedefaultmidpunct}
{\mcitedefaultendpunct}{\mcitedefaultseppunct}\relax
\EndOfBibitem
\bibitem[Sun \latin{et~al.}(2007)Sun, DuFort, Daniel, Murali, Chen, Gopinath,
  Stein, De, Rotello, Holzenburg, \latin{et~al.} others]{sun2007core}
Sun,~J.; DuFort,~C.; Daniel,~M.-C.; Murali,~A.; Chen,~C.; Gopinath,~K.;
  Stein,~B.; De,~M.; Rotello,~V.~M.; Holzenburg,~A., \latin{et~al.}
  Core-Controlled Polymorphism in Virus-like Particles. \emph{Proc. Nat. Acad.
  Sci.} \textbf{2007}, \emph{104}, 1354--1359\relax
\mciteBstWouldAddEndPuncttrue
\mciteSetBstMidEndSepPunct{\mcitedefaultmidpunct}
{\mcitedefaultendpunct}{\mcitedefaultseppunct}\relax
\EndOfBibitem
\bibitem[Huang \latin{et~al.}(2007)Huang, Bronstein, Retrum, Dufort, Tsvetkova,
  Aniagyei, Stein, Stucky, McKenna, Remmes, \latin{et~al.}
  others]{huang2007self}
Huang,~X.; Bronstein,~L.~M.; Retrum,~J.; Dufort,~C.; Tsvetkova,~I.;
  Aniagyei,~S.; Stein,~B.; Stucky,~G.; McKenna,~B.; Remmes,~N., \latin{et~al.}
  Self-Assembled Virus-like Particles with Magnetic Cores. \emph{Nano Letters}
  \textbf{2007}, \emph{7}, 2407--2416\relax
\mciteBstWouldAddEndPuncttrue
\mciteSetBstMidEndSepPunct{\mcitedefaultmidpunct}
{\mcitedefaultendpunct}{\mcitedefaultseppunct}\relax
\EndOfBibitem
\bibitem[Dixit \latin{et~al.}(2006)Dixit, Goicochea, Daniel, Murali, Bronstein,
  De, Stein, Rotello, Kao, and Dragnea]{dixit2006quantum}
Dixit,~S.~K.; Goicochea,~N.~L.; Daniel,~M.-C.; Murali,~A.; Bronstein,~L.;
  De,~M.; Stein,~B.; Rotello,~V.~M.; Kao,~C.~C.; Dragnea,~B. Quantum Dot
  Encapsulation in Viral Capsids. \emph{Nano Letters} \textbf{2006}, \emph{6},
  1993--1999\relax
\mciteBstWouldAddEndPuncttrue
\mciteSetBstMidEndSepPunct{\mcitedefaultmidpunct}
{\mcitedefaultendpunct}{\mcitedefaultseppunct}\relax
\EndOfBibitem
\bibitem[Kuenzle \latin{et~al.}(2018)Kuenzle, Mangler, Lach, and
  Beck]{Kuenzle2018}
Kuenzle,~M.; Mangler,~J.; Lach,~M.; Beck,~T. Peptide-directed encapsulation of
  inorganic nanoparticles into protein containers. \emph{NANOSCALE}
  \textbf{2018}, \emph{10}, 22917--22926\relax
\mciteBstWouldAddEndPuncttrue
\mciteSetBstMidEndSepPunct{\mcitedefaultmidpunct}
{\mcitedefaultendpunct}{\mcitedefaultseppunct}\relax
\EndOfBibitem
\bibitem[McPherson(2005)]{mcpherson2005micelle}
McPherson,~A. Micelle Formation and Crystallization as Paradigms for Virus
  Assembly. \emph{Bioessays} \textbf{2005}, \emph{27}, 447--458\relax
\mciteBstWouldAddEndPuncttrue
\mciteSetBstMidEndSepPunct{\mcitedefaultmidpunct}
{\mcitedefaultendpunct}{\mcitedefaultseppunct}\relax
\EndOfBibitem
\bibitem[Devkota \latin{et~al.}(2009)Devkota, Petrov, Lemieux, Boz, Tang,
  Schneemann, Johnson, and Harvey]{devkota2009structural}
Devkota,~B.; Petrov,~A.~S.; Lemieux,~S.; Boz,~M.~B.; Tang,~L.; Schneemann,~A.;
  Johnson,~J.~E.; Harvey,~S.~C. Structural and Electrostatic Characterization
  of Pariacoto Virus: Implications for Viral Assembly. \emph{Biopolymers}
  \textbf{2009}, \emph{91}, 530--538\relax
\mciteBstWouldAddEndPuncttrue
\mciteSetBstMidEndSepPunct{\mcitedefaultmidpunct}
{\mcitedefaultendpunct}{\mcitedefaultseppunct}\relax
\EndOfBibitem
\bibitem[Hagan(2008)]{hagan2008controlling}
Hagan,~M.~F. Controlling Viral Capsid Assembly with Templating. \emph{Phys.
  Rev. E} \textbf{2008}, \emph{77}, 051904\relax
\mciteBstWouldAddEndPuncttrue
\mciteSetBstMidEndSepPunct{\mcitedefaultmidpunct}
{\mcitedefaultendpunct}{\mcitedefaultseppunct}\relax
\EndOfBibitem
\bibitem[Elrad and Hagan(2010)Elrad, and Hagan]{elrad2010encapsulation}
Elrad,~O.~M.; Hagan,~M.~F. Encapsulation of a Polymer by an Icosahedral Virus.
  \emph{Phys. Biol.} \textbf{2010}, \emph{7}, 045003\relax
\mciteBstWouldAddEndPuncttrue
\mciteSetBstMidEndSepPunct{\mcitedefaultmidpunct}
{\mcitedefaultendpunct}{\mcitedefaultseppunct}\relax
\EndOfBibitem
\bibitem[Panahandeh \latin{et~al.}(2020)Panahandeh, Li, Marichal, Leite~Rubim,
  Tresset, and Zandi]{panahandeh2020virus}
Panahandeh,~S.; Li,~S.; Marichal,~L.; Leite~Rubim,~R.; Tresset,~G.; Zandi,~R.
  How a Virus Circumvents Energy Barriers to Form Symmetric Shells. \emph{ACS
  Nano} \textbf{2020}, \emph{14}, 3170--3180\relax
\mciteBstWouldAddEndPuncttrue
\mciteSetBstMidEndSepPunct{\mcitedefaultmidpunct}
{\mcitedefaultendpunct}{\mcitedefaultseppunct}\relax
\EndOfBibitem
\bibitem[Malyutin and Dragnea(2013)Malyutin, and Dragnea]{Malyutin2013}
Malyutin,~A.~G.; Dragnea,~B. Budding Pathway in the Templated Assembly of
  Viruslike Particles. \emph{J. Phys. Chem. B} \textbf{2013}, \emph{117},
  10730--10736\relax
\mciteBstWouldAddEndPuncttrue
\mciteSetBstMidEndSepPunct{\mcitedefaultmidpunct}
{\mcitedefaultendpunct}{\mcitedefaultseppunct}\relax
\EndOfBibitem
\bibitem[Chevreuil \latin{et~al.}(2018)Chevreuil, Law-Hine, Chen, Bressanelli,
  Combet, Constantin, Degrouard, M{\"{o}}ller, Zeghal, and
  Tresset]{Chevreuil2018}
Chevreuil,~M.; Law-Hine,~D.; Chen,~J.; Bressanelli,~S.; Combet,~S.;
  Constantin,~D.; Degrouard,~J.; M{\"{o}}ller,~J.; Zeghal,~M.; Tresset,~G.
  Nonequilibrium Self-Assembly Dynamics of Icosahedral Viral Capsids Packaging
  Genome or Polyelectrolyte. \emph{Nat. Commun.} \textbf{2018}, \emph{9},
  3071\relax
\mciteBstWouldAddEndPuncttrue
\mciteSetBstMidEndSepPunct{\mcitedefaultmidpunct}
{\mcitedefaultendpunct}{\mcitedefaultseppunct}\relax
\EndOfBibitem
\bibitem[Garmann \latin{et~al.}(2019)Garmann, Goldfain, and
  Manoharan]{garmann2019measurements}
Garmann,~R.~F.; Goldfain,~A.~M.; Manoharan,~V.~N. Measurements of the
  Self-Assembly Kinetics of Individual Viral Capsids around their RNA Genome.
  \emph{Proc. Natl. Acad. Sci.} \textbf{2019}, \emph{116}, 22485--22490\relax
\mciteBstWouldAddEndPuncttrue
\mciteSetBstMidEndSepPunct{\mcitedefaultmidpunct}
{\mcitedefaultendpunct}{\mcitedefaultseppunct}\relax
\EndOfBibitem
\bibitem[Li \latin{et~al.}(2024)Li, Tresset, and Zandi]{li2024elastomer}
Li,~S.; Tresset,~G.; Zandi,~R. Switchable Conformation in Protein Subunits:
  Unveiling Assembly Dynamics of Icosahedral Viruses (preprint). \textbf{2024},
  \relax
\mciteBstWouldAddEndPunctfalse
\mciteSetBstMidEndSepPunct{\mcitedefaultmidpunct}
{}{\mcitedefaultseppunct}\relax
\EndOfBibitem
\bibitem[Chang \latin{et~al.}(2008)Chang, Knobler, Gelbart, and
  Mason]{Chang2008}
Chang,~C.~B.; Knobler,~C.~M.; Gelbart,~W.~M.; Mason,~T.~G. Curvature Dependence
  of Viral Protein Structures on Encapsidated Nanoemulsion Droplets. \emph{ACS
  Nano} \textbf{2008}, \emph{2}, 281--286\relax
\mciteBstWouldAddEndPuncttrue
\mciteSetBstMidEndSepPunct{\mcitedefaultmidpunct}
{\mcitedefaultendpunct}{\mcitedefaultseppunct}\relax
\EndOfBibitem
\bibitem[Hu \latin{et~al.}(2008)Hu, Zandi, Anavitarte, Knobler, and
  Gelbart]{Hu2008}
Hu,~Y.; Zandi,~R.; Anavitarte,~A.; Knobler,~C.~M.; Gelbart,~W.~M. Packaging of
  a Polymer by a Viral Capsid: The Interplay between Polymer Length and Capsid
  Size. \emph{Biophys. J.} \textbf{2008}, \emph{94}, 1428--36\relax
\mciteBstWouldAddEndPuncttrue
\mciteSetBstMidEndSepPunct{\mcitedefaultmidpunct}
{\mcitedefaultendpunct}{\mcitedefaultseppunct}\relax
\EndOfBibitem
\bibitem[Tanaka \latin{et~al.}(2008)Tanaka, Kerfeld, Sawaya, Cai, Heinhorst,
  Cannon, and Yeates]{Tanaka2008}
Tanaka,~S.; Kerfeld,~C.~A.; Sawaya,~M.~R.; Cai,~F.; Heinhorst,~S.;
  Cannon,~G.~C.; Yeates,~T.~O. Atomic-Level Models of the Bacterial Carboxysome
  Shell. \emph{Science} \textbf{2008}, \emph{319}, 1083--1086\relax
\mciteBstWouldAddEndPuncttrue
\mciteSetBstMidEndSepPunct{\mcitedefaultmidpunct}
{\mcitedefaultendpunct}{\mcitedefaultseppunct}\relax
\EndOfBibitem
\bibitem[Perlmutter \latin{et~al.}(2016)Perlmutter, Mohajerani, and
  Hagan]{Perlmutter2016}
Perlmutter,~J.~D.; Mohajerani,~F.; Hagan,~M.~F. Many-Molecule Encapsulation by
  an Icosahedral Shell. \emph{eLife} \textbf{2016}, \emph{5}, e14078\relax
\mciteBstWouldAddEndPuncttrue
\mciteSetBstMidEndSepPunct{\mcitedefaultmidpunct}
{\mcitedefaultendpunct}{\mcitedefaultseppunct}\relax
\EndOfBibitem
\bibitem[Rotskoff and Geissler(2018)Rotskoff, and Geissler]{Rotskoff2018}
Rotskoff,~G.~M.; Geissler,~P.~L. Robust Non-Equilibrium Pathways to
  Microcompartment Assembly. \emph{Proc. Nat. Acad. Sci.} \textbf{2018},
  \emph{115}, 6341--6346\relax
\mciteBstWouldAddEndPuncttrue
\mciteSetBstMidEndSepPunct{\mcitedefaultmidpunct}
{\mcitedefaultendpunct}{\mcitedefaultseppunct}\relax
\EndOfBibitem
\bibitem[Beck \latin{et~al.}(2015)Beck, Tetter, K{\"u}nzle, and
  Hilvert]{Beck2015}
Beck,~T.; Tetter,~S.; K{\"u}nzle,~M.; Hilvert,~D. Construction of
  Matryoshka-Type Structures from Supercharged Protein Nanocages. \emph{Angew.
  Chem., Int. Ed.} \textbf{2015}, \emph{54}, 937--940\relax
\mciteBstWouldAddEndPuncttrue
\mciteSetBstMidEndSepPunct{\mcitedefaultmidpunct}
{\mcitedefaultendpunct}{\mcitedefaultseppunct}\relax
\EndOfBibitem
\bibitem[Moerman \latin{et~al.}(2016)Moerman, van~der Schoot, and
  Kegel]{Moerman2016}
Moerman,~P.; van~der Schoot,~P.; Kegel,~W. Kinetics Versus Thermodynamics in
  Virus Capsid Polymorphism. \emph{J. Phys. Chem. B} \textbf{2016}, \emph{120},
  6003--6009\relax
\mciteBstWouldAddEndPuncttrue
\mciteSetBstMidEndSepPunct{\mcitedefaultmidpunct}
{\mcitedefaultendpunct}{\mcitedefaultseppunct}\relax
\EndOfBibitem
\bibitem[van~der Holst \latin{et~al.}(2018)van~der Holst, Kegel, Zandi, and
  van~der Schoot]{VanderHolst2018}
van~der Holst,~B.; Kegel,~W.~K.; Zandi,~R.; van~der Schoot,~P. The different
  faces of mass action in virus assembly. \emph{J Biol. Phys.} \textbf{2018},
  \emph{44}, 163--179\relax
\mciteBstWouldAddEndPuncttrue
\mciteSetBstMidEndSepPunct{\mcitedefaultmidpunct}
{\mcitedefaultendpunct}{\mcitedefaultseppunct}\relax
\EndOfBibitem
\bibitem[Kao and Sivakumaran(2000)Kao, and Sivakumaran]{kao2000brome}
Kao,~C.~C.; Sivakumaran,~K. Brome Mosaic Virus, Good for an RNA Virologist’s
  Basic Needs. \emph{Mol. Plant Pathol.} \textbf{2000}, \emph{1}, 91--97\relax
\mciteBstWouldAddEndPuncttrue
\mciteSetBstMidEndSepPunct{\mcitedefaultmidpunct}
{\mcitedefaultendpunct}{\mcitedefaultseppunct}\relax
\EndOfBibitem
\bibitem[Kao \latin{et~al.}(2011)Kao, Ni, Hema, Huang, and Dragnea]{Kao2011}
Kao,~C.~C.; Ni,~P.; Hema,~M.; Huang,~X.; Dragnea,~B. The Coat Protein Leads the
  Way: An Update on Basic and Applied Studies with the Brome Mosaic Virus Coat
  Protein. \emph{Mol. Plant Pathol.} \textbf{2011}, \emph{12}, 403--412\relax
\mciteBstWouldAddEndPuncttrue
\mciteSetBstMidEndSepPunct{\mcitedefaultmidpunct}
{\mcitedefaultendpunct}{\mcitedefaultseppunct}\relax
\EndOfBibitem
\bibitem[Casjens(1985)]{Casjens1985}
Casjens,~S. \emph{Virus Structure and Assembly.}; Jones {\{}{\&}{\}} Bartlett
  Pub, 1985\relax
\mciteBstWouldAddEndPuncttrue
\mciteSetBstMidEndSepPunct{\mcitedefaultmidpunct}
{\mcitedefaultendpunct}{\mcitedefaultseppunct}\relax
\EndOfBibitem
\bibitem[Lucas \latin{et~al.}(2002)Lucas, Larson, and
  McPherson]{lucas2002crystallographic}
Lucas,~R.~W.; Larson,~S.~B.; McPherson,~A. The Crystallographic Structure of
  Brome Mosaic Virus. \emph{J. Mol. Biol.} \textbf{2002}, \emph{317},
  95--108\relax
\mciteBstWouldAddEndPuncttrue
\mciteSetBstMidEndSepPunct{\mcitedefaultmidpunct}
{\mcitedefaultendpunct}{\mcitedefaultseppunct}\relax
\EndOfBibitem
\bibitem[Kegel and van~der Schoot(2004)Kegel, and van~der
  Schoot]{kegel2004competing}
Kegel,~W.~K.; van~der Schoot,~P. Competing Hydrophobic and Screened-Coulomb
  Interactions in Hepatitis B Virus Capsid Assembly. \emph{Biophys. J.}
  \textbf{2004}, \emph{86}, 3905--3913\relax
\mciteBstWouldAddEndPuncttrue
\mciteSetBstMidEndSepPunct{\mcitedefaultmidpunct}
{\mcitedefaultendpunct}{\mcitedefaultseppunct}\relax
\EndOfBibitem
\bibitem[Ceres and Zlotnick(2002)Ceres, and Zlotnick]{ceres2002weak}
Ceres,~P.; Zlotnick,~A. Weak Protein-Protein Interactions are Sufficient to
  Drive Assembly of Hepatitis B Virus Capsids. \emph{Biochem.} \textbf{2002},
  \emph{41}, 11525--11531\relax
\mciteBstWouldAddEndPuncttrue
\mciteSetBstMidEndSepPunct{\mcitedefaultmidpunct}
{\mcitedefaultendpunct}{\mcitedefaultseppunct}\relax
\EndOfBibitem
\bibitem[Zandi \latin{et~al.}(2020)Zandi, Dragnea, Travesset, and
  Podgornik]{zandi2020virus}
Zandi,~R.; Dragnea,~B.; Travesset,~A.; Podgornik,~R. On Virus Growth and Form.
  \emph{Physics Reports} \textbf{2020}, \emph{847}, 1--102\relax
\mciteBstWouldAddEndPuncttrue
\mciteSetBstMidEndSepPunct{\mcitedefaultmidpunct}
{\mcitedefaultendpunct}{\mcitedefaultseppunct}\relax
\EndOfBibitem
\bibitem[Belyi and Muthukumar(2006)Belyi, and
  Muthukumar]{belyi2006electrostatic}
Belyi,~V.~A.; Muthukumar,~M. Electrostatic Origin of the Genome Packing in
  Viruses. \emph{Proc. Nat. Acad. Sci.} \textbf{2006}, \emph{103},
  17174--17178\relax
\mciteBstWouldAddEndPuncttrue
\mciteSetBstMidEndSepPunct{\mcitedefaultmidpunct}
{\mcitedefaultendpunct}{\mcitedefaultseppunct}\relax
\EndOfBibitem
\bibitem[Hu \latin{et~al.}(2008)Hu, Zhang, and Shklovskii]{hu2008electrostatic}
Hu,~T.; Zhang,~R.; Shklovskii,~B. Electrostatic Theory of Viral Self-Assembly.
  \emph{Phys.} \textbf{2008}, \emph{387}, 3059--3064\relax
\mciteBstWouldAddEndPuncttrue
\mciteSetBstMidEndSepPunct{\mcitedefaultmidpunct}
{\mcitedefaultendpunct}{\mcitedefaultseppunct}\relax
\EndOfBibitem
\bibitem[van~der Schoot and Bruinsma(2005)van~der Schoot, and
  Bruinsma]{van2005electrostatics}
van~der Schoot,~P.; Bruinsma,~R. Electrostatics and the Assembly of an RNA
  Virus. \emph{Phys. Rev. E.} \textbf{2005}, \emph{71}, 061928\relax
\mciteBstWouldAddEndPuncttrue
\mciteSetBstMidEndSepPunct{\mcitedefaultmidpunct}
{\mcitedefaultendpunct}{\mcitedefaultseppunct}\relax
\EndOfBibitem
\bibitem[Krol \latin{et~al.}(1999)Krol, Olson, Tate, Johnson, Baker, and
  Ahlquist]{krol1999rna}
Krol,~M.~A.; Olson,~N.~H.; Tate,~J.; Johnson,~J.~E.; Baker,~T.~S.; Ahlquist,~P.
  RNA-Controlled Polymorphism in the in vivo Assembly of 180-Subunit and
  120-Subunit Virions from a Single Capsid Protein. \emph{Proc. Nat. Acad.
  Sci.} \textbf{1999}, \emph{96}, 13650--13655\relax
\mciteBstWouldAddEndPuncttrue
\mciteSetBstMidEndSepPunct{\mcitedefaultmidpunct}
{\mcitedefaultendpunct}{\mcitedefaultseppunct}\relax
\EndOfBibitem
\bibitem[Dragnea \latin{et~al.}(2003)Dragnea, Chen, Kwak, Stein, and
  Kao]{dragnea2003gold}
Dragnea,~B.; Chen,~C.; Kwak,~E.-S.; Stein,~B.; Kao,~C.~C. Gold Nanoparticles as
  Spectroscopic Enhancers for in vitro Studies on Single Viruses. \emph{J. Am.
  Chem. Soc.} \textbf{2003}, \emph{125}, 6374--6375\relax
\mciteBstWouldAddEndPuncttrue
\mciteSetBstMidEndSepPunct{\mcitedefaultmidpunct}
{\mcitedefaultendpunct}{\mcitedefaultseppunct}\relax
\EndOfBibitem
\bibitem[Zeng \latin{et~al.}(2018)Zeng, Rodriguez~L{\'a}zaro, Tsvetkova, Hagan,
  and Dragnea]{Zeng2018}
Zeng,~C.; Rodriguez~L{\'a}zaro,~G.; Tsvetkova,~I.~B.; Hagan,~M.~F.; Dragnea,~B.
  Defects and Chirality in the Nanoparticle-Directed Assembly of
  Spherocylindrical Shells of Virus Coat Proteins. \emph{ACS Nano}
  \textbf{2018}, \emph{12}, 5323--5332\relax
\mciteBstWouldAddEndPuncttrue
\mciteSetBstMidEndSepPunct{\mcitedefaultmidpunct}
{\mcitedefaultendpunct}{\mcitedefaultseppunct}\relax
\EndOfBibitem
\bibitem[Zhao \latin{et~al.}(1995)Zhao, Fox, Olson, Baker, and
  Young]{zhao1995vitro}
Zhao,~X.; Fox,~J.~M.; Olson,~N.~H.; Baker,~T.~S.; Young,~M.~J. In vitro
  Assembly of Cowpea Chlorotic Mottle Virus from Coat Protein Expressed in
  Escherichia coli and in vitro-transcribed Viral cDNA. \emph{Virology}
  \textbf{1995}, \emph{207}, 486--494\relax
\mciteBstWouldAddEndPuncttrue
\mciteSetBstMidEndSepPunct{\mcitedefaultmidpunct}
{\mcitedefaultendpunct}{\mcitedefaultseppunct}\relax
\EndOfBibitem
\bibitem[Caspar and Klug(1962)Caspar, and Klug]{Caspar1962}
Caspar,~D. L.~D.; Klug,~A. Physical Principles in Construction of Regular
  Viruses. \emph{Cold Spring Harb. Symp. Quant. Biol.} \textbf{1962},
  \emph{27}, 1--24\relax
\mciteBstWouldAddEndPuncttrue
\mciteSetBstMidEndSepPunct{\mcitedefaultmidpunct}
{\mcitedefaultendpunct}{\mcitedefaultseppunct}\relax
\EndOfBibitem
\bibitem[Daniel \latin{et~al.}(2010)Daniel, Tsvetkova, Quinkert, Murali, De,
  Rotello, Kao, and Dragnea]{daniel2010role}
Daniel,~M.-C.; Tsvetkova,~I.~B.; Quinkert,~Z.~T.; Murali,~A.; De,~M.;
  Rotello,~V.~M.; Kao,~C.~C.; Dragnea,~B. Role of Surface Charge Density in
  Nanoparticle-Templated Assembly of Bromovirus Protein Cages. \emph{ACS Nano}
  \textbf{2010}, \emph{4}, 3853--3860\relax
\mciteBstWouldAddEndPuncttrue
\mciteSetBstMidEndSepPunct{\mcitedefaultmidpunct}
{\mcitedefaultendpunct}{\mcitedefaultseppunct}\relax
\EndOfBibitem
\bibitem[He \latin{et~al.}(2013)He, Porterfield, van~der Schoot, Zlotnick, and
  Dragnea]{He2013}
He,~L.; Porterfield,~Z.; van~der Schoot,~P.; Zlotnick,~A.; Dragnea,~B.
  Hepatitis Virus Capsid Polymorph Stability Depends on Encapsulated Cargo
  Size. \emph{ACS Nano} \textbf{2013}, \emph{7}, 8447--8454\relax
\mciteBstWouldAddEndPuncttrue
\mciteSetBstMidEndSepPunct{\mcitedefaultmidpunct}
{\mcitedefaultendpunct}{\mcitedefaultseppunct}\relax
\EndOfBibitem
\bibitem[Kusters \latin{et~al.}(2015)Kusters, Lin, Zandi, Tsvetkova, Dragnea,
  and van~der Schoot]{kusters2015role}
Kusters,~R.; Lin,~H.-K.; Zandi,~R.; Tsvetkova,~I.; Dragnea,~B.; van~der
  Schoot,~P. Role of Charge Regulation and Size Polydispersity in Nanoparticle
  Encapsulation by Viral Coat Proteins. \emph{J. Phys. Chem. B} \textbf{2015},
  \emph{119}, 1869--1880\relax
\mciteBstWouldAddEndPuncttrue
\mciteSetBstMidEndSepPunct{\mcitedefaultmidpunct}
{\mcitedefaultendpunct}{\mcitedefaultseppunct}\relax
\EndOfBibitem
\bibitem[Xia \latin{et~al.}(2012)Xia, Yang, Wang, Zheng, Li, Chen, and
  Xia]{Xia2012}
Xia,~X.; Yang,~M.; Wang,~Y.; Zheng,~Y.; Li,~Q.; Chen,~J.; Xia,~Y. Quantifying
  the Coverage Density of Poly(ethylene glycol) Chains on the Surface of Gold
  Nanostructures. \emph{ACS Nano} \textbf{2012}, \emph{6}, 512--522\relax
\mciteBstWouldAddEndPuncttrue
\mciteSetBstMidEndSepPunct{\mcitedefaultmidpunct}
{\mcitedefaultendpunct}{\mcitedefaultseppunct}\relax
\EndOfBibitem
\bibitem[Garmann \latin{et~al.}(2016)Garmann, Comas-Garcia, Knobler, and
  Gelbart]{garmann2016physical}
Garmann,~R.~F.; Comas-Garcia,~M.; Knobler,~C.~M.; Gelbart,~W.~M. Physical
  Principles in the Self-Assembly of a Simple Spherical Virus. \emph{Acc. Chem.
  Res.} \textbf{2016}, \emph{49}, 48--55\relax
\mciteBstWouldAddEndPuncttrue
\mciteSetBstMidEndSepPunct{\mcitedefaultmidpunct}
{\mcitedefaultendpunct}{\mcitedefaultseppunct}\relax
\EndOfBibitem
\bibitem[Lin \latin{et~al.}(2012)Lin, van~der Schoot, and Zandi]{lin2012impact}
Lin,~H.-K.; van~der Schoot,~P.; Zandi,~R. Impact of Charge Variation on the
  Encapsulation of Nanoparticles by Virus Coat Proteins. \emph{Phys. Biol.}
  \textbf{2012}, \emph{9}, 066004\relax
\mciteBstWouldAddEndPuncttrue
\mciteSetBstMidEndSepPunct{\mcitedefaultmidpunct}
{\mcitedefaultendpunct}{\mcitedefaultseppunct}\relax
\EndOfBibitem
\bibitem[van~der Schoot \latin{et~al.}(2025)van~der Schoot, Zandi, and
  Dragnea]{vdSchoot2025}
van~der Schoot,~P.; Zandi,~R.; Dragnea,~B. Mass Action and the Encapsulation of
  Fragmented Cargo by Virus Coat Proteins. \emph{manuscript in preparation}
  \textbf{2025}, \relax
\mciteBstWouldAddEndPunctfalse
\mciteSetBstMidEndSepPunct{\mcitedefaultmidpunct}
{}{\mcitedefaultseppunct}\relax
\EndOfBibitem
\bibitem[Gonzalez~Solveyra and Szleifer(2016)Gonzalez~Solveyra, and
  Szleifer]{Solveyra2016}
Gonzalez~Solveyra,~E.; Szleifer,~I. What is the Role of Curvature on the
  Properties of Nanomaterials for Biomedical Applications? \emph{Wiley
  Interdiscip. Rev.} \textbf{2016}, \emph{8}, 334--354\relax
\mciteBstWouldAddEndPuncttrue
\mciteSetBstMidEndSepPunct{\mcitedefaultmidpunct}
{\mcitedefaultendpunct}{\mcitedefaultseppunct}\relax
\EndOfBibitem
\bibitem[Timmermans \latin{et~al.}(2022)Timmermans, Ramezani, Montalvo, Nguyen,
  van~der Schoot, van Hest, and Zandi]{Timmermans2022}
Timmermans,~S. B. P.~E.; Ramezani,~A.; Montalvo,~T.; Nguyen,~M.; van~der
  Schoot,~P.; van Hest,~J. C.~M.; Zandi,~R. The Dynamics of Virus-like Capsid
  Assembly and Disassembly. \emph{J. Am. Chem. Soc.} \textbf{2022}, \emph{144},
  12608--12612\relax
\mciteBstWouldAddEndPuncttrue
\mciteSetBstMidEndSepPunct{\mcitedefaultmidpunct}
{\mcitedefaultendpunct}{\mcitedefaultseppunct}\relax
\EndOfBibitem
\bibitem[Tsai and Lee(2011)Tsai, and Lee]{Tsai2011}
Tsai,~P.; Lee,~E. Gel Electrophoresis in Suspensions of Charged Spherical
  Particles. \emph{Soft Matter} \textbf{2011}, \emph{7}, 5789--5798\relax
\mciteBstWouldAddEndPuncttrue
\mciteSetBstMidEndSepPunct{\mcitedefaultmidpunct}
{\mcitedefaultendpunct}{\mcitedefaultseppunct}\relax
\EndOfBibitem
\bibitem[Henry(1931)]{henry1931cataphoresis}
Henry,~D.~C. The Cataphoresis of Suspended Particles. Part I.—The Equation of
  Cataphoresis. \emph{Proc. R. Soc. London} \textbf{1931}, \emph{133},
  106--129\relax
\mciteBstWouldAddEndPuncttrue
\mciteSetBstMidEndSepPunct{\mcitedefaultmidpunct}
{\mcitedefaultendpunct}{\mcitedefaultseppunct}\relax
\EndOfBibitem
\bibitem[Wang \latin{et~al.}(2011)Wang, Nap, Lagzi, Kowalczyk, Han, Grzybowski,
  and Szleifer]{wang2011and}
Wang,~D.; Nap,~R.~J.; Lagzi,~I.; Kowalczyk,~B.; Han,~S.; Grzybowski,~B.~A.;
  Szleifer,~I. How and Why Nanoparticle’s Curvature Regulates the Apparent p
  K a of the Coating Ligands. \emph{J. Am. Chem. Soc.} \textbf{2011},
  \emph{133}, 2192--2197\relax
\mciteBstWouldAddEndPuncttrue
\mciteSetBstMidEndSepPunct{\mcitedefaultmidpunct}
{\mcitedefaultendpunct}{\mcitedefaultseppunct}\relax
\EndOfBibitem
\bibitem[Nayak \latin{et~al.}(2024)Nayak, Zhang, Bu, Ocko, Travesset, Vaknin,
  Mallapragada, and Wang]{nayak2024ionic}
Nayak,~B.~P.; Zhang,~H.; Bu,~W.; Ocko,~B.~M.; Travesset,~A.; Vaknin,~D.;
  Mallapragada,~S.~K.; Wang,~W. Ionic-like Superlattices by Charged
  Nanoparticles: A Step Toward Photonics Applications. \emph{ACS Appl. Nano
  Mater.} \textbf{2024}, \emph{7}, 3220--3228\relax
\mciteBstWouldAddEndPuncttrue
\mciteSetBstMidEndSepPunct{\mcitedefaultmidpunct}
{\mcitedefaultendpunct}{\mcitedefaultseppunct}\relax
\EndOfBibitem
\bibitem[Ostuni \latin{et~al.}(2001)Ostuni, Chapman, Holmlin, Takayama, and
  Whitesides]{ostuni2001survey}
Ostuni,~E.; Chapman,~R.~G.; Holmlin,~R.~E.; Takayama,~S.; Whitesides,~G.~M. A
  Survey of Structure-Property Relationships of Surfaces that Resist the
  Adsorption of Protein. \emph{Langmuir} \textbf{2001}, \emph{17},
  5605--5620\relax
\mciteBstWouldAddEndPuncttrue
\mciteSetBstMidEndSepPunct{\mcitedefaultmidpunct}
{\mcitedefaultendpunct}{\mcitedefaultseppunct}\relax
\EndOfBibitem
\bibitem[You \latin{et~al.}(2006)You, Verma, and Rotello]{you2006engineering}
You,~C.-C.; Verma,~A.; Rotello,~V.~M. Engineering the
  Nanoparticle--Biomacromolecule Interface. \emph{Soft Matter} \textbf{2006},
  \emph{2}, 190--204\relax
\mciteBstWouldAddEndPuncttrue
\mciteSetBstMidEndSepPunct{\mcitedefaultmidpunct}
{\mcitedefaultendpunct}{\mcitedefaultseppunct}\relax
\EndOfBibitem
\bibitem[Meng \latin{et~al.}(2014)Meng, Paulose, Nelson, and
  Manoharan]{meng2014elastic}
Meng,~G.; Paulose,~J.; Nelson,~D.~R.; Manoharan,~V.~N. Elastic Instability of a
  Crystal Growing on a Curved Surface. \emph{Science} \textbf{2014},
  \emph{343}, 634--637\relax
\mciteBstWouldAddEndPuncttrue
\mciteSetBstMidEndSepPunct{\mcitedefaultmidpunct}
{\mcitedefaultendpunct}{\mcitedefaultseppunct}\relax
\EndOfBibitem
\bibitem[Paquay \latin{et~al.}(2016)Paquay, Kusumaatmaja, Wales, Zandi, and
  van~der Schoot]{paquay2016energetically}
Paquay,~S.; Kusumaatmaja,~H.; Wales,~D.~J.; Zandi,~R.; van~der Schoot,~P.
  Energetically Favoured Defects in Dense Packings of Particles on Spherical
  Surfaces. \emph{Soft Matter} \textbf{2016}, \emph{12}, 5708--5717\relax
\mciteBstWouldAddEndPuncttrue
\mciteSetBstMidEndSepPunct{\mcitedefaultmidpunct}
{\mcitedefaultendpunct}{\mcitedefaultseppunct}\relax
\EndOfBibitem
\bibitem[Luque \latin{et~al.}(2012)Luque, Reguera, Morozov, Rudnick, and
  Bruinsma]{Luque2012}
Luque,~A.; Reguera,~D.; Morozov,~A.; Rudnick,~J.; Bruinsma,~R. Physics of Shell
  Assembly: Line Tension, Hole Implosion, and Closure Catastrophe. \emph{J.
  Chem. Phys.} \textbf{2012}, \emph{136}, 184507\relax
\mciteBstWouldAddEndPuncttrue
\mciteSetBstMidEndSepPunct{\mcitedefaultmidpunct}
{\mcitedefaultendpunct}{\mcitedefaultseppunct}\relax
\EndOfBibitem
\bibitem[Morozov \latin{et~al.}(2009)Morozov, Bruinsma, and
  Rudnick]{morozov2009assembly}
Morozov,~A.; Bruinsma,~R.; Rudnick,~J. Assembly of Viruses and the Pseudo Law
  of Mass Action. \emph{Biophys. J.} \textbf{2009}, \emph{96}, 419a--420a\relax
\mciteBstWouldAddEndPuncttrue
\mciteSetBstMidEndSepPunct{\mcitedefaultmidpunct}
{\mcitedefaultendpunct}{\mcitedefaultseppunct}\relax
\EndOfBibitem
\bibitem[Thuman-Commike \latin{et~al.}(1998)Thuman-Commike, Greene, Malinski,
  King, and Chiu]{thuman1998role}
Thuman-Commike,~P.~A.; Greene,~B.; Malinski,~J.~A.; King,~J.; Chiu,~W. Role of
  the Scaffolding Protein in P22 Procapsid Size Determination suggested by T= 4
  and T= 7 Procapsid Structures. \emph{Biophys. J.} \textbf{1998}, \emph{74},
  559--568\relax
\mciteBstWouldAddEndPuncttrue
\mciteSetBstMidEndSepPunct{\mcitedefaultmidpunct}
{\mcitedefaultendpunct}{\mcitedefaultseppunct}\relax
\EndOfBibitem
\bibitem[Earnshaw and King(1978)Earnshaw, and King]{earnshaw1978structure}
Earnshaw,~W.; King,~J. Structure of Phage P22 Coat Protein Aggregates Formed in
  the Absence of the Scaffolding Protein. \emph{J. of Mol. Bio.} \textbf{1978},
  \emph{126}, 721--747\relax
\mciteBstWouldAddEndPuncttrue
\mciteSetBstMidEndSepPunct{\mcitedefaultmidpunct}
{\mcitedefaultendpunct}{\mcitedefaultseppunct}\relax
\EndOfBibitem
\bibitem[Salunke \latin{et~al.}(1989)Salunke, Caspar, and
  Garcea]{salunke1989polymorphism}
Salunke,~D.; Caspar,~D.; Garcea,~R. Polymorphism in the Assembly of
  Polyomavirus Capsid Protein VP1. \emph{Biophys. J.} \textbf{1989}, \emph{56},
  887--900\relax
\mciteBstWouldAddEndPuncttrue
\mciteSetBstMidEndSepPunct{\mcitedefaultmidpunct}
{\mcitedefaultendpunct}{\mcitedefaultseppunct}\relax
\EndOfBibitem
\bibitem[Erickson and Pantaloni(1981)Erickson, and Pantaloni]{erickson1981role}
Erickson,~H.; Pantaloni,~D. The Role of Subunit Entropy in Cooperative
  Assembly: Nucleation of Microtubules and Other Two-Dimensional Polymers.
  \emph{Biophys. J.} \textbf{1981}, \emph{34}, 293--309\relax
\mciteBstWouldAddEndPuncttrue
\mciteSetBstMidEndSepPunct{\mcitedefaultmidpunct}
{\mcitedefaultendpunct}{\mcitedefaultseppunct}\relax
\EndOfBibitem
\bibitem[Maassen \latin{et~al.}(2018)Maassen, de~Ruiter, Lindhoud, and
  Cornelissen]{maassen2018oligonucleotide}
Maassen,~S.~J.; de~Ruiter,~M.~V.; Lindhoud,~S.; Cornelissen,~J.~J.
  Oligonucleotide Length-Dependent Formation of Virus-like Particles.
  \emph{Chem. - Eur. J.} \textbf{2018}, \emph{24}, 7456--7463\relax
\mciteBstWouldAddEndPuncttrue
\mciteSetBstMidEndSepPunct{\mcitedefaultmidpunct}
{\mcitedefaultendpunct}{\mcitedefaultseppunct}\relax
\EndOfBibitem
\bibitem[Gopinath(2007)]{Gopinath2007}
Gopinath,~C.~C.,~K.;~Kao Replication-Independent Long-Distance Trafficking by
  Viral RNAs in Nicotiana benthamiana. \emph{Plant Cell.} \textbf{2007},
  \emph{19}, 1179--91\relax
\mciteBstWouldAddEndPuncttrue
\mciteSetBstMidEndSepPunct{\mcitedefaultmidpunct}
{\mcitedefaultendpunct}{\mcitedefaultseppunct}\relax
\EndOfBibitem
\bibitem[Peng \latin{et~al.}(2008)Peng, Lee, Wang, Yin, Dai, and Sun]{peng2008}
Peng,~S.; Lee,~Y.; Wang,~C.; Yin,~H.; Dai,~S.; Sun,~S. A facile synthesis of
  monodisperse Au nanoparticles and their catalysis of CO oxidation. \emph{Nano
  Research.} \textbf{2008}, \emph{1}, 229--234\relax
\mciteBstWouldAddEndPuncttrue
\mciteSetBstMidEndSepPunct{\mcitedefaultmidpunct}
{\mcitedefaultendpunct}{\mcitedefaultseppunct}\relax
\EndOfBibitem
\bibitem[Vieweger \latin{et~al.}(2018)Vieweger, Tsvetkova, and
  Dragnea]{Vieweger2018}
Vieweger,~S.~E.; Tsvetkova,~I.~B.; Dragnea,~B.~G. In Vitro Assembly of
  Virus-Derived Designer Shells Around Inorganic Nanoparticles. \emph{Methods
  Mol. Biol.} \textbf{2018}, \emph{1776}, 279--294\relax
\mciteBstWouldAddEndPuncttrue
\mciteSetBstMidEndSepPunct{\mcitedefaultmidpunct}
{\mcitedefaultendpunct}{\mcitedefaultseppunct}\relax
\EndOfBibitem
\bibitem[Mastronarde and Held(2017)Mastronarde, and
  Held]{Mastronarde2017AutomatedTS}
Mastronarde,~D.~N.; Held,~S.~R. Automated Tilt Series Alignment and Tomographic
  Reconstruction in IMOD. \emph{J. Struct. Biol.} \textbf{2017}, \emph{197 2},
  102--113\relax
\mciteBstWouldAddEndPuncttrue
\mciteSetBstMidEndSepPunct{\mcitedefaultmidpunct}
{\mcitedefaultendpunct}{\mcitedefaultseppunct}\relax
\EndOfBibitem
\end{mcitethebibliography}

\end{document}